\documentclass[letterpaper,twocolumn,10pt]{article}
\usepackage{usenix}
\usepackage[available]{usenixbadges}

\usepackage{xcolor}
\usepackage{enumitem}
\usepackage{amsmath}
\usepackage{booktabs}
\usepackage{graphicx}
\usepackage{verbatim}
\usepackage{xspace}
\usepackage{multirow}
\usepackage{threeparttable} 
\usepackage[bb=boondox]{mathalfa}
\usepackage{pifont}
\usepackage{marvosym}
\usepackage{xurl}
\usepackage{arydshln}
\usepackage{makecell}
\usepackage[framemethod=TikZ]{mdframed}
\usepackage{lipsum}
\usepackage{algorithm}
\usepackage{cancel}
\usepackage{algpseudocode}
\usepackage{utfsym}
\usepackage{MnSymbol}
\usepackage{subcaption}
\usepackage{colortbl}
\usepackage{textcomp}
\usepackage{multicol}
\usepackage{longtable}
\newcommand*\circled[1]{\tikz[baseline=(char.base)]{
            \node[shape=circle,draw,inner sep=1pt] (char) {#1};}}

\makeatletter
\newcommand{\ALC@uniqueautorefname}{Line}
\makeatother

\usepackage{etoolbox}
\makeatletter
\patchcmd{\hyper@makecurrent}{%
    \ifx\Hy@param\Hy@chapterstring
        \let\Hy@param\Hy@chapapp
    \fi
}{%
    \iftoggle{inappendix}{
        \@checkappendixparam{chapter}%
        \@checkappendixparam{section}%
        \@checkappendixparam{subsection}%
        \@checkappendixparam{subsubsection}%
        \@checkappendixparam{paragraph}%
        \@checkappendixparam{subparagraph}%
    }{}%
}{}{\errmessage{failed to patch}}

\newcommand*{\@checkappendixparam}[1]{%
    \def\@checkappendixparamtmp{#1}%
    \ifx\Hy@param\@checkappendixparamtmp
        \let\Hy@param\Hy@appendixstring
    \fi
}
\makeatletter

\newtoggle{inappendix}
\togglefalse{inappendix}

\apptocmd{\appendix}{\toggletrue{inappendix}}{}{\errmessage{failed to patch}}

\newcommand{\customref}[1]{\hyperref[#1]{C\ref*{#1}}}

\setlist[itemize]{topsep=0pt, itemsep=0pt, parsep=0pt, partopsep=0pt}
\setlist[itemize]{leftmargin=8pt}

\usepackage{filecontents}

\begin{document}

\newcommand\CQ[1]{\textbf{\textcolor{purple}{CQ: #1}}}
\newcommand\Jiayi[1]{\textbf{\textcolor{orange}{Jiayi: #1}}}

\newcommand{\sys}{\textsc{XGuardian}\xspace}
\newcommand{\inspector}{\textsc{Inspector}\xspace}
\newcommand{\explainer}{\textsc{Explainer}\xspace}

\date{}

\title{\Large \bf \sys: Towards Generalized, Explainable and More Effective Server-side Anti-cheat in First-Person Shooter Games}

\author{
{\rm Jiayi Zhang}\\
The University of Hong Kong\\
brucejiayi@connect.hku.hk
\and
{\rm  Chenxin Sun}\\
The University of Hong Kong\\
roniny@connect.hku.hk
\and
{\rm Chenxiong Qian}\\
The University of Hong Kong\\
cqian@cs.hku.hk
} 

\maketitle
\begin{abstract}
Aim-assist cheats are the most prevalent and infamous form of cheating in First-Person Shooter (FPS) games, which help cheaters illegally reveal the opponent's location and auto-aim and shoot, and thereby pose significant threats to the game industry. 
Although a considerable research effort has been made to automatically detect aim-assist cheats, existing works suffer from unreliable frameworks, limited generalizability, high overhead, low detection performance, and a lack of explainability of detection results.
In this paper, we propose \sys, a server-side generalized and explainable system for detecting aim-assist cheats to overcome these limitations. 
It requires only two raw data inputs, pitch and yaw, which are all FPS games' must-haves, to construct novel temporal features and describe aim trajectories, which are essential for distinguishing cheaters and normal players.
\sys is evaluated with the latest mainstream FPS game CS2, and validates its generalizability with two different games.
It achieves high detection performance and low overhead compared to prior works across different games with real-world and large-scale datasets, demonstrating wide generalizability and high effectiveness.
It is able to justify its predictions and thereby shorten the manual review latency.
We make \sys and our datasets publicly available.
\end{abstract}
\section{Introduction}
\label{sec:introduction}
The global online gaming market is projected to grow to approximately  \$250.2 billion by the end of 2025, with PC gaming leading as the fastest-growing segment, anticipated to expand by 8.1\% in 2025 \cite{MobileMarketingReads2025}. 
Within this landscape, First-Person Shooter (FPS) games hold a significant 20.9\% market share and are expected to maintain a compound annual growth rate of 8.1\% from 2025 to 2033 \cite{wepc2023,DataHorizzon2025}.
Despite their popularity, FPS games are particularly vulnerable to cheating, as developers of malicious software exploit their competitive and multiplayer nature for monetary gain \cite{hawk2024}.
This pervasive issue not only undermines fair play but also poses substantial economic and reputational risks to the gaming industry \cite{ExponentialEra2024}.
%

%
Among the various forms of cheating, aim-assist cheats are the most prevalent and notorious.
These cheats enable players to illegally reveal opponents' locations, auto-aim and shoot, providing unfair advantages even to seasoned or professional players \cite{choi2023botscreen}.
This has spurred significant interest from both academia and industry in developing effective anti-cheat solutions. 
Numerous approaches have been proposed, leveraging statistical features and machine learning techniques to identify cheaters \cite{galli2011unreal,yu2012statistical,liu2017detecting, alayed2013ml,yu2012heuristic}, while others have employed advanced time-series or ensemble models to detect temporal anomalies \cite{choi2023botscreen, hawk2024}.
However, existing solutions are hampered by several critical limitations, which we categorize as \textbf{L1} through \textbf{L4}.

\definecolor{lightred}{HTML}{cb2424}
\definecolor{lightgreen}{HTML}{1e8b7a}

\begin{table*}[htbp]
\centering
\caption{ML-based anti-cheat detection works comparison.}
\vspace{-0.5em}
\label{tab:piror_work_review}
\begin{threeparttable}
\resizebox{\textwidth}{!}{%
\begin{tabular}{@{}cllcccccccc@{}}
\toprule
 &  &  &  & \multicolumn{3}{c}{\textbf{Generalizability}} &  & \multicolumn{3}{c}{\textbf{Real-world   Functionality}} \\ \cmidrule(lr){5-7} \cmidrule(l){9-11} 
\multirow{-2}{*}{\textbf{Side}} & \multirow{-2}{*}{\textbf{Method}} & \multirow{-2}{*}{\textbf{Work(s)}} & \multirow{-2}{*}{\textbf{Availability}\tnote{1}} & \textbf{Data Collection} & \textbf{Feature}\tnote{2} & \textbf{Evaluated}\tnote{3} & \multirow{-2}{*}{\textbf{Interpretability}} & \textbf{Dataset} & \textbf{System Constraint} & \textbf{Performance} \\ \midrule
\multirow{2}{*}{\textbf{Client}} & Unsupervised Angle Anomaly Detection & \cite{choi2023botscreen} & {\textbf{\textcolor{lightgreen}{T}, \textcolor{lightgreen}{C}, \textcolor{lightgreen}{D}}} & $\usym{2613}$ & $\bcancel{\checkmark}$ & $\usym{2613}$ & $\usym{2613}$ & SSL\tnote{4} & TEE (Intel SGX) & Unknown\tnote{5} \\
 & Statistical Classification & \cite{yu2012statistical,yu2012heuristic} & {\textbf{\textcolor{lightgreen}{T}, \textcolor{lightred}{C}, \textcolor{lightred}{D}}} & $\usym{2613}$ & $\checkmark$ & $\usym{2613}$ & $\usym{2613}$ & SSL & Concurrent overhead & Unknown \\ \midrule
\multirow{3}{*}{\textbf{Server}} & Statistical Classification & \cite{galli2011unreal,liu2017detecting, alayed2013ml} & {\textbf{\textcolor{lightgreen}{T}, \textcolor{lightred}{C}, \textcolor{lightred}{D}}} & $\usym{2613}$ & $\checkmark$ & $\usym{2613}$ & $\usym{2613}$ & SSL & Concurrent overhead & Unknown \\
 & Multi-view Ensemble Learning & \cite{hawk2024} & {\textbf{\textcolor{lightgreen}{T}, \textcolor{lightgreen}{C}, \textcolor{lightgreen}{D}}} & $\checkmark$ & $\bcancel{\checkmark}$ & $\usym{2613}$ & $\usym{2613}$ & Real-world & None & Low \\
\cmidrule(l){2-11}
 & Temporal Aim Trajectory Classification & \textbf{\sys} & {\textbf{\textcolor{lightgreen}{T}, \textcolor{lightgreen}{C}, \textcolor{lightgreen}{D}}} & $\checkmark$ & $\checkmark$ & $\checkmark$ & $\checkmark$ & \text{Real-world} & \text{None} & \text{High} \\ \bottomrule
\end{tabular}%
}
\newline
\raggedright \footnotesize
\textsuperscript{1}\textbf{Availability}: Public availability of technical details (T), source code (C), and datasets (D), \textcolor{lightred}{red} and \textcolor{lightgreen}{green} indicate unavailable and available;
\textsuperscript{2}\textbf{Feature}: $\bcancel{\checkmark}$ denotes part or all of the features are not generalized;
\textsuperscript{3}\textbf{Evaluated}: If the method's generalizability is evaluated through experiment(s);
\textsuperscript{4}\textbf{SSL}: Simulated, small-scale, low complexity;
\textsuperscript{5}\textbf{Unknown}: Have not been evaluated. 
\end{threeparttable}
\end{table*}

\noindent\textbf{(L1) Client-side Dependency and Resource Constraints.} Many recent solutions rely on client-side data collection and cheat detection \cite{yu2012statistical,choi2023botscreen,yu2012heuristic}.
However, this approach is fraught with challenges, including the ease of bypassing anti-cheat mechanisms, risks of memory tampering, hardware limitations, and system overhead in highly concurrent environments.
Cheaters can circumvent most current anti-cheat systems by using obfuscated cheats and higher privileges \cite{9774028}, or by directly accessing memory to tamper with detection models or processes \cite{choi2023botscreen}.
Although the latest design, \textsc{BotScreen} \cite{choi2023botscreen}, addresses some of these issues using Intel SGX, a Trusted Execution Environment (TEE), SGX was deprecated in 2021 \cite{bleepingcomputer2024, intelcommunity2024}, and its limited compatibility and practicality hinder widespread industrial adoption.
Moreover, client-side solutions must operate in highly concurrent environments, where the client handles game logic, rendering, voice chat, and anti-cheat data extraction and analysis simultaneously. 
This consumption of scarce client-side resources increases hardware requirements, complicates optimization, and degrades the overall gaming experience.

\noindent\textbf{(L2) Limited Real-world Performance.} The real-world efficacy of most previous server-side \cite{galli2011unreal,liu2017detecting, alayed2013ml} and client-side \cite{yu2012statistical,choi2023botscreen,yu2012heuristic} solutions remains uncertain, as they have only been tested on small-scale, simulated datasets.
This lack of real-world validation could lead to performance degradation or failure in complex scenarios. 
While the latest server-side solution \cite{hawk2024} is evaluated on a large-scale, real-world dataset and avoids client-side limitations (detailed in \autoref{sec:server-side_anti-cheat_background}), its prediction performance remains relatively low due to the complexity of the data.
Additionally, increasing community grievances about FPS anti-cheat systems indicate low recall rates in current industrial solutions \cite{zleague_warzone_anticheat, pushtotalk_anticheat, algshack2024, hawk2024}.

\noindent\textbf{(L3) Limited Generalizability.} Existing solutions often lack generalizability.
For instance, the feature design of the most advanced server-side solution \cite{hawk2024} is tailored to a specific FPS subgenre, tactical shooters \cite{tactical_shooter}. 
It requires features like economy- and grenade-related features, making it difficult to generalize across different sub-genres of FPS games, as its authors acknowledge.
While some statistical features in other works, such as playtime and winning rates \cite{han2015online}, are generalizable, they are not directly relevant to aim-assist cheats \cite{choi2023botscreen}, reducing prediction accuracy.
Other features, like the angle between the line of sight and a target used in the latest client-side solution \cite{choi2023botscreen}, require specific engine modifications for different games, as such data may not exist in log or replay files  (introduced in \autoref{sec:replay_system}), limiting their applicability. 
%
%
%

\noindent\textbf{(L4) Lack of Explainability.} A critical drawback in all prior solutions is the lack of interpretability.
A ban requires manual verification, which involves reviewers reviewing hours of the suspect's gameplay replays.
This process is expensive and time-consuming for the industry.
None of the existing anti-cheat works can explain how or why a prediction is made, which is essential for any practical anti-cheat system to pinpoint the exact suspicious segments so that it can justify the ban and shorten the review time.

In this paper, we introduce \textbf{Ex}plainable \textbf{Guardian} (\sys) to address the aforementioned limitations.
Deployed on the server side, \sys avoids the inherent limitations of client-side solutions and leverages replay files as the data source to perform detection asynchronously during server idle periods, minimizing overhead \textbf{(L1)}.
\sys achieves state-of-the-art performance on a large, complex real-world dataset \textbf{(L2)} and is the first work to demonstrate wide generalizability by validating its effectiveness on large datasets from different sub-genres of games \textbf{(L3)}. 
\sys addresses the explainability bottleneck by instantly pinpointing the exact suspicious frames and features within a match. 
This drastically reduces manual review time, making our system a powerful and practical tool for game moderators. 
We also conduct case studies to illustrate \sys's interpretability \textbf{(L4)}.
\autoref{tab:piror_work_review} lists the high-level comparison between \sys and prior works.

The key innovation of \sys lies in its ability to standardize and generalize the representation of player aiming trajectories across all FPS games, while also providing an interpretable analysis of these trajectories.
\sys is generalized by requiring only two raw inputs, pitch and yaw (introduced in \autoref{sec:pitch_yaw}), which are must-haves in all FPS games. 
%
%
%
We discover three fundamental aspects, velocity, acceleration, and angular change, to effectively represent a player's aiming trajectory in a generalized manner, making it applicable to detecting cheaters. 
%
%
\sys employs a GRU-CNN model to effectively classify time-series features, enhances the explainability framework SHAP (introduced in \autoref{sec:xai}), and designs the corresponding visualizations to provide interpretable detection results.

%
We implemented and evaluated \sys on the well-known FPS game Counter-Strike 2 (CS2), using a large-scale real-world dataset from an active platform with millions of users. 
The dataset is two orders of magnitude larger than that of the most advanced client-side solution \cite{choi2023botscreen} and matches the data volume of the leading server-side solution \cite{hawk2024}, both analyzing approximately 3,000 matches.
Specifically, \sys analyzes 3,069,216 aiming operations across 31,971 aim trajectories from 5,486 players in real-world matches. 
It achieves up to 90.7\% recall while maintaining a false positive rate (FPR) of 4.1\%. 
Compared to previous anti-cheat methods, \sys improves detection accuracy by 12.5\% over the state-of-the-art method.
Importantly, to the best of our knowledge, \sys is the first work in anti-cheat detection to prove its generalizability through a large-scale experiment on different games in real practice.
We demonstrate that \sys's performance remains consistent across different sub-genres of games on different platforms (PC and mobile).
We empirically show that \sys remains somewhat robust under adversarial attacks.
\sys also achieves interpretability by illustrating each elimination trajectory explanation from different feature aspects.
Additionally, \sys incurs marginal overhead on the server.
Our contributions are as follows:
%
%
\begin{itemize}[topsep=6pt, itemsep=2pt, parsep=0pt, partopsep=0pt]
\item We propose novel, generalized, and effective features, a time-series GRU-CNN model, and an explainable detection framework for identifying aim-assist cheats in games.
\item Our approach demonstrates strong performance and evaluated with the real-world data.
\item We open-source \sys and release two new datasets (CS2, Farlight84) in \nameref{sec:openscience}.
\end{itemize}
\section{Background}
\label{sec:background}

\subsection{Server-side Anti-cheat}
\label{sec:server-side_anti-cheat_background}
Server-side anti-cheat systems are defined by their architectural design to collect, process, and analyze gameplay data exclusively on secure server infrastructure. 
While client-side telemetry remains inherently vulnerable to tampering (as discussed in \textbf{L1}), server-side detection offers a critical advantage: cheaters must ultimately manifest their advantages through observable server-authoritative actions to impact gameplay outcomes \cite{hawk2024,choi2023botscreen}. 
By aggregating final-state player inputs and game-state snapshots, these systems establish a ground-truth dataset resistant to client-side manipulation.

Earlier server-side anti-cheat systems \cite{alayed2013ml,galli2011unreal,liu2017detecting,yeung2006detecting} relied on simplistic threshold-based heuristics or single-model classifiers using coarse-grained metrics, e.g., hit-accuracy, win-rate distributions.
However, the proliferation of replay systems in modern FPS games (see \autoref{sec:replay_system}) has enabled richer server-side data pipelines, characterized by modularity and loose coupling between game engines and analytical frameworks.
The state-of-the-art anti-cheat design, \textsc{Hawk} \cite{hawk2024}, leverage this paradigm shift through multi-view feature engineering, combining spatial-temporal behavioral patterns with deep learning architectures to detect sophisticated cheats. 
Nevertheless, two critical limitations persist: (a) existing systems lack interpretable decision mechanisms, and (b) detection models exhibit poor generalizability across all FPS games.

\subsection{Aim-Assist Cheats in FPS Games}
\label{sec:cheat_type}
Modern cheating tools have evolved into integrated platforms that combine multiple exploit modalities.
For example, in most modern FPS games, \textbf{aim-assist} cheats commonly bundle aimbots (auto-aiming) with wallhacks (see-through walls), creating compounded unfair advantages \cite{SecureCheats2025}. 
%
%
In terms of the form, aim-assist cheats have evolved from pure aimbots (direct reticle manipulation to target centroids) to triggerbots (automated firing conditioned on crosshair overlap) and computer-vision-based aimbots (real-time object detection via screen scraping)~\cite{invisibilitycloak}.
%
%
%
%
%
Moreover, cheaters dynamically adjust sophistication levels of using aim-assist cheats across two axes: (1) functional granularity (e.g., targeting specific body part vs. full-body locks) and (2) stealth tradeoffs (blatant play vs. behavioral mimicry of legitimate players).
%
%
%
%
Thus, anti-cheat systems have shifted from signature-based detection of individual cheat types to behavioral analysis of cheat categories \cite{dfsec,dfbans}.
%
%
As aim-assist cheats constitute a significant portion of all cheating behaviors for FPS games \cite{hawk2024,choi2023botscreen,blackmirror2020}, our work focuses on addressing the aim-assist cheats.
Our evaluations in \autoref{sec:evaluation} addresses this through multi-dimensional sophistication modeling across large-scale real-world datasets.

\subsection{Replay Systems and Replay Files}
\label{sec:replay_system}
Modern FPS engines incorporate replay systems that serialize match state transitions into game-agnostic replay files (e.g., CS2's \textit{.dem} format). 
%
These files provide a complete spatiotemporal record of player inputs, game events, and entity states, decoupled from runtime engine dependencies. 
%
%
\sys operates on server-authoritative post-match records, which provide three practical benefits for behavioral cheat analysis: (1) resistance to client-side manipulation, (2) avoiding introducing real-time overhead constraints, (3) access to a unified global game state for camera-independent behavior reconstruction.
Considering server-side solutions require direct cooperation from game developers, which is a resource rarely accessible to the broader research community, replay files as an anti-cheat data source can be further leveraged for offline detection methods, as replay files can be parsed and analyzed independently of the game server to extract relevant data.
Broader researchers can likewise use \sys to assess the presence of cheaters in a given match.

\subsection{SHAP Explainers}
\label{sec:xai}
Model interpretability is critical for anti-cheat systems for adversarial robustness and fairness auditing. 
%
SHAP \cite{shap2017,treeshap} addresses this by quantifying feature contributions through coalitional game theory. 
It establishes a connection between optimal credit allocation and local explanations by calculating classic Shapley values, which explain the ML model's output by distributing the difference between the actual prediction and the model's baseline prediction across all input features.
%
%
%
A positive Shapley value indicates that the feature significantly contributes to increasing the prediction and vice versa.
\mbox{\textit{GradientExplainer}} and \mbox{\textit{TreeExplainer}} are SHAP's implementations used for interpreting complex ML models and are customized and used in \sys for the decision explanation. 
\mbox{\textit{GradientExplainer}} is suited for deep learning models and leverages model gradients to explain how input features impact the prediction. 
\mbox{\textit{TreeExplainer}} is designed for tree-based models to compute optimized Shapley values.
%
\section{Overview}
\label{sec:overview}

\subsection{Threat Model}
\label{sec:threat_model}
To reflect realistic anti-cheat deployment scenarios, we assume that adversaries utilize aim-assist cheats to gain an unfair competitive advantage in FPS games.
We model the specific type, configuration, and level of sophistication of these cheats as black boxes, focusing on their observable behavioral impact. While we cannot exhaustively evaluate every possible cheat variant beyond our dataset, this assumption captures the diversity and unpredictability of real-world cheating behaviors, where attackers may use commercially available or custom-developed tools with varying degrees of subtlety or aggression.
For instance, one adversary may employ a wallhack from vendor X, deliberately disabling visual enhancements such as enemy model highlighting while enabling minimap position overlays to mimic legitimate play and avoid detection.
Conversely, another may use a combination of a wallhack and a hard-lock aimbot from vendor Y, bypassing visual obfuscations (e.g., flashbangs or smoke grenades in CS2) and engaging in overt, easily noticeable cheating.
Our approach aims to detect a wide spectrum of aim-assist behavior, from stealthy to blatant, without relying on prior knowledge of the cheat’s implementation details.

\begin{figure}[htbp]
\centering
\includegraphics[width=0.47\textwidth]{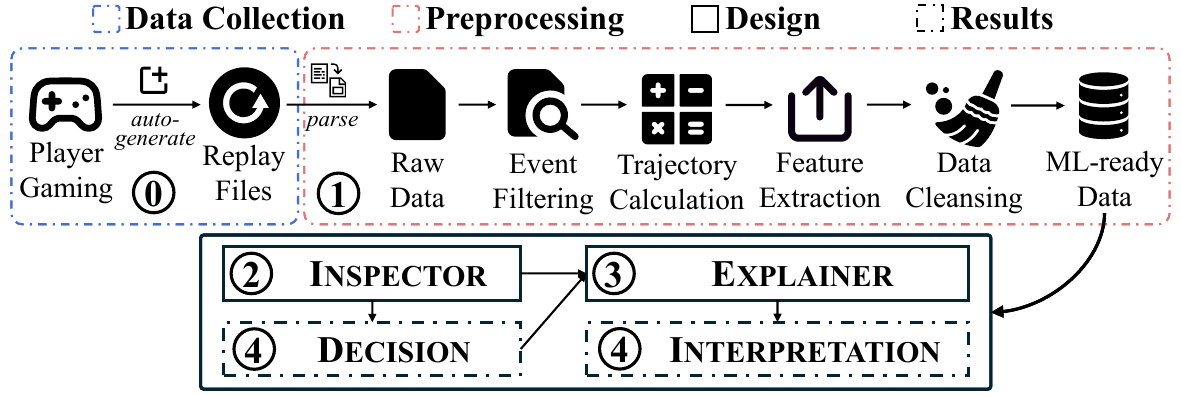}
\caption{Workflow of \sys.}
\label{fig:workflow}
\end{figure}

\subsection{Workflow}
\label{sec:workflow}
\autoref{fig:workflow} presents the overall workflow of \sys, which operates in four main stages following gameplay.

\noindent\circled{0} \textbf{Data Collection}. 
Players engage in matches as usual, during which the game engine automatically records gameplay into replay files upon match completion.
These replay files are collected from our partner platform, which also provides ground truth labels indicating whether the player has cheated.

\noindent\circled{1} \textbf{Preprocessing}.
We begin by parsing raw time-series data from the replay files (referred to \textit{demos} in CS2) into structured tabular form.
For each player, we identify the \textit{ticks}, the smallest time unit in the game, corresponding to \textit{elimination events} (conduct a kill on an opponent), and extract a fixed-length time window surrounding each event.
Within this window, we compute the player's crosshair trajectory in 2D screen space, representing their real-world mouse-controlled aiming path.
From these trajectories, we extract a comprehensive set of features as detailed in \autoref{sec:features}, thereby constructing a dataset for model training and evaluation.

\noindent\circled{2} \textbf{Inspection}. The preprocessed samples that correspond to an elimination event are then passed to \inspector.
\inspector employs an ensemble of temporal classification models to analyze the samples and detect anomalous patterns indicative of aim-assist behavior.
Its core responsibility is to dynamically assess whether suspicious behavior occurs during specific eliminations within a match.
Further architectural and implementation details are provided in \autoref{sec:inspector}.

\noindent\circled{3} \textbf{Explanation}.
After classification, \inspector's outputs are combined with the training data to serve as background context for \explainer.
\explainer quantifies the influence of input features on each classification result, offering interpretable insights into the detected aiming behaviors and the broader in-match performance.
A detailed discussion of the explanation module is provided in \autoref{sec:explainer}.

\noindent\circled{4} \textbf{Decision and Explainability Output}.
For each player in each match, \sys produces a binary decision indicating whether cheating was detected.
Each decision is accompanied by a visualized explanation at feature and match level respectively, providing transparency into how and why the system arrived at its conclusion.

\section{\sys}
\label{sec:design}
\subsection{Trajectory Extraction and Feature Design}
\label{sec:features}
\subsubsection{Dimensions of movement in FPS games}
\label{sec:pitch_yaw}
\begin{figure}[htbp]
\centering
\begin{subfigure}[htbp]{0.17\textwidth}
    \centering
    \includegraphics[width=\textwidth]{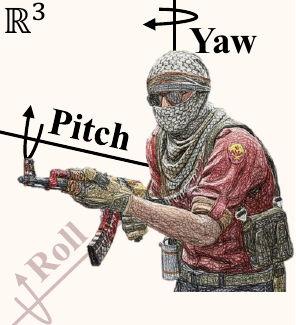}
    \caption{Pitch and yaw}
    \label{fig:pitch_yaw}
\end{subfigure}
\vfill
\begin{subfigure}[htbp]{0.17\textwidth}
    \centering
    \includegraphics[width=\textwidth]{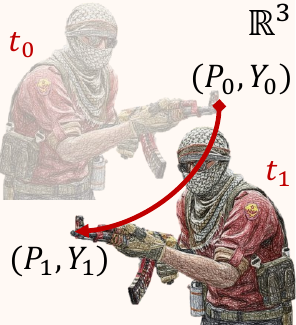}
    \caption{Game world}
    \label{fig:3d_feature}
\end{subfigure}
\hfill
\begin{subfigure}[htbp]{0.17\textwidth}
    \centering
    \includegraphics[width=\textwidth]{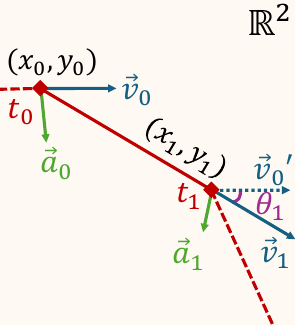}
    \caption{Monitor screen}
    \label{fig:2d_feature}
\end{subfigure}
\caption{Pitch and yaw, the dimensions of movement (a); aiming trajectory extraction in 3D game world (b) and feature constructions in 2D monitor screen (c).}
\label{fig:feature_vis}
\end{figure}

\autoref{fig:pitch_yaw} illustrates the three rotational dimensions commonly used to describe player view orientation in 3D first-person shooter (FPS) environments: \textit{pitch}, \textit{yaw}, and \textit{roll}.
All these angles are defined in the game engine's world coordinate system.
Among these, pitch and yaw are the primary angular measurements used to represent a player's aiming direction and camera orientation.
\textbf{Pitch} refers to the vertical angle of the player's view. It captures the up-and-down rotation of the camera, with values typically ranging from -90$^{\circ}$ to 90$^{\circ}$.
A pitch of -90$^{\circ}$ indicates that the player is looking directly downward, 0$^{\circ}$ corresponds to a level forward gaze, and 90$^{\circ}$ indicates a view directly upward.
\textbf{Yaw} represents the horizontal angle of the player's view, measuring the left-to-right rotation.
Yaw typically ranges from -180$^{\circ}$ to 180$^{\circ}$. A yaw of 0$^{\circ}$ indicates a forward-facing direction, -90$^{\circ}$ indicates a leftward view, 90$^{\circ}$ indicates a rightward view, and ±180$^{\circ}$ corresponds to looking directly behind.
Roll describes rotation around the forward axis of the camera, i.e., the tilt of the view to the left or right.
While roll is a valid component of 3D orientation, it is largely irrelevant in the context of FPS gameplay, where camera tilt does not impact aiming or navigation.
Consequently, \sys focuses exclusively on pitch and yaw for modeling player aiming behavior.

\subsubsection{Extraction scope}
\label{sec:extraction_scope}
\begin{figure}[htbp]
\centering
\includegraphics[width=0.47\textwidth]{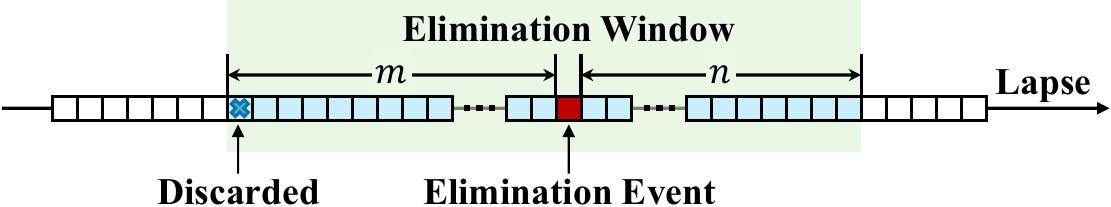}
\caption{Elimination window selection.}
\label{fig:elimination_window}
\end{figure}

Throughout an entire match, players spend most of their time outside of combat, meaning their aiming trajectories often contain little useful information and may instead introduce noise and unnecessary computational overhead.
To address this, we define an \textit{elimination window}, which is created when a player triggers an \textit{elimination event}. 
Specifically, we locate the corresponding \textit{tick} of the event and extract data from $m$ \textit{ticks} before and $n$ \textit{ticks} after it.
The window size is $(m-1)+n+1=m+n$.
These selected \textit{ticks} form an elimination window and serve as individual data samples.
\autoref{fig:elimination_window} illustrates the elimination window selection process.
To prevent adversarial attacks, \sys focuses on analyzing elimination, an unavoidable and attack-resistant game-critical event, rather than other events like firing that cheaters may easily manipulate (e.g., spam random shots to introduce noise and poison the model).

\subsubsection{3D aiming trajectory extraction and 2D mapping}
\label{sec:trajectory_extraction}
In all FPS games, players use input peripherals, typically a mouse or controller, to control their in-game avatar’s aim within a 3D virtual environment.
This aiming behavior is governed by two angular measurements: \textit{pitch} and \textit{yaw}, which represent vertical and horizontal orientations.
The player’s aiming trajectory in the 3D world can thus be expressed as a sequence of pitch and yaw values over time. As illustrated in \autoref{fig:3d_feature}, the red curve denotes a trajectory beginning at time tick $t_0$ with angles $(P_0, Y_0)$ and ending at $t_1$ with angles $(P_1, Y_1)$, where $P$ and $Y$ denote pitch and yaw, respectively.

To enable effective visualization and downstream analysis, these 3D world-space angles must be mapped into 2D screen coordinates, simulating how the player’s crosshair moves across the display.
The raw data extracted from the \textit{demo} files includes pitch and yaw values, which are converted to 2D coordinates using the transformation process detailed in \autoref{alg:py2xy}.
Specifically, yaw values, which range from [-180$^{\circ}$, 180$^{\circ}$], are first normalized to [0$^{\circ}$, 360$^{\circ}$] by adding 180$^{\circ}$ and then scaled according to the screen width.
Similarly, pitch values, ranging from [-90$^{\circ}$, 90$^{\circ}$], are normalized to [0$^{\circ}$, 180$^{\circ}$] by adding 90$^{\circ}$ and scaled to the screen height.
To conform to the screen coordinate system, where the origin is at the top-left corner, the Y-axis is then inverted.

\begin{algorithm}
\caption{Convert Pitch and Yaw to 2D Coordinates}
\label{alg:py2xy}
\begin{algorithmic}[1]
\Function{PitchYawToXY}{pitch, yaw, width, height}
    \State $x \gets \left(\frac{yaw + 180}{360}\right) \times width$
    \State $y \gets \left(\frac{pitch + 90}{180}\right) \times height$
    \State $y \gets height - y$
    \State \Return $(x, y)$
\EndFunction
\end{algorithmic}
\end{algorithm}

The conversion result is shown in \autoref{fig:2d_feature}, where the solid red line represents the corresponding 2D crosshair trajectory mapped from the 3D trajectory shown in \autoref{fig:3d_feature}, spanning from $t_0$ to $t_1$.
Importantly, pitch and yaw are standardized components across all FPS games, enabling \sys to generalize across a broad range of titles in this genre.

\begin{figure}[htbp]
\centering
\begin{subfigure}[htbp]{0.47\textwidth}
    \centering
    \includegraphics[width=\textwidth]{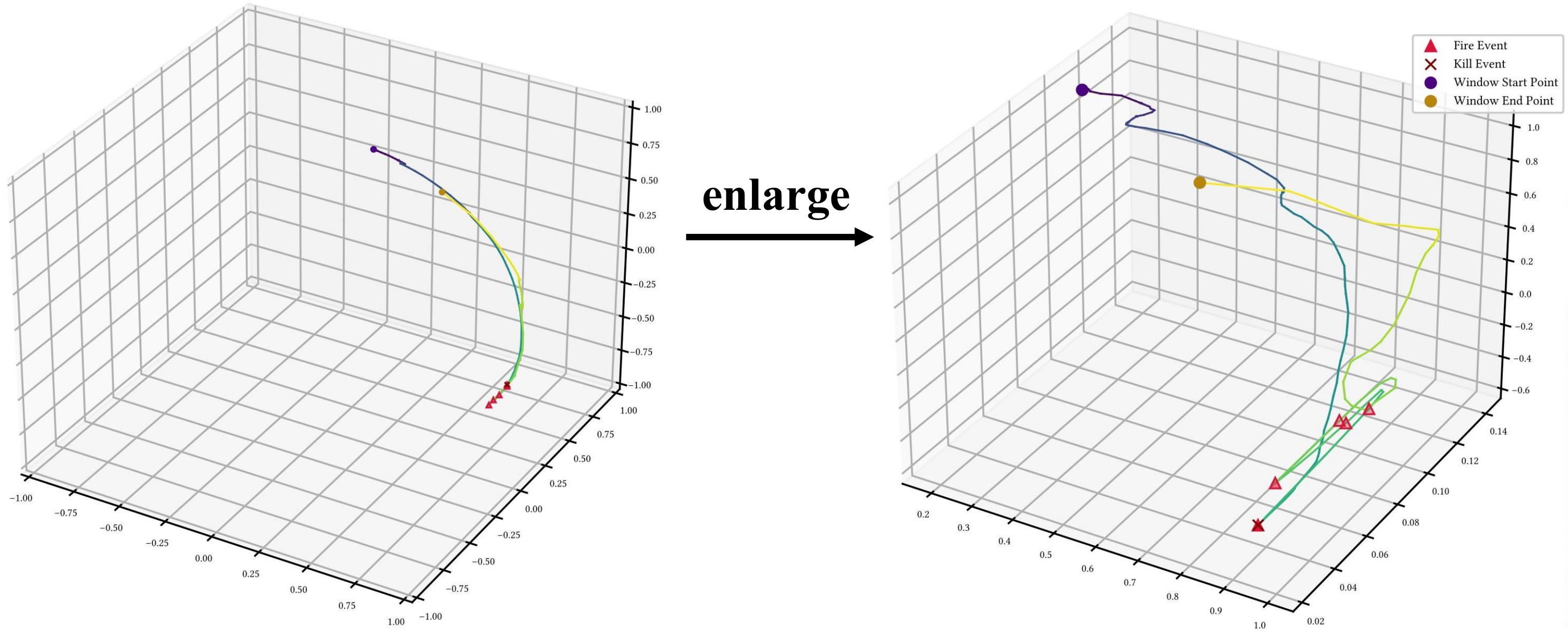}
    \caption{Game world aiming trajectory}
    \label{fig:3d_vis_trajectory}
\end{subfigure}
\begin{subfigure}[htbp]{0.4\textwidth}
    \centering
    \includegraphics[width=\textwidth]{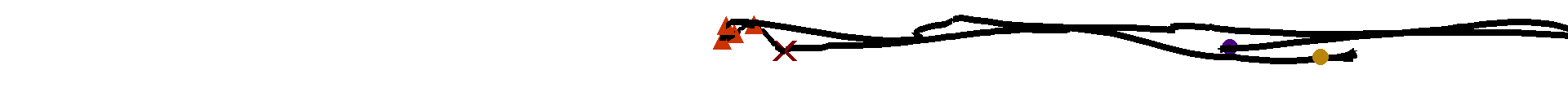}
    \caption{Monitor screen aiming trajectory}
    \label{fig:2d_vis_trajectory}
\end{subfigure}
\caption{The example of 3D trajectory (a) and 2D screen trajectory mapping (b). The enlarged 3D trajectory (a-right) uses different measurement units for each coordinate axis for better illustration. 2D trajectory visualization is mapped onto a 1920$\times$1080 resolution, and the coordinates of the elimination event are relocated to the center of the screen.}
\label{fig:vis_trajectory}
\end{figure}

We further visualize the player’s aiming trajectory in the 3D game environment in \autoref{fig:3d_vis_trajectory}.
It reflects how the player adjusts their view direction, via pitch and yaw, to target opponents.
The trajectory begins at the purple point and ends at the yellow point, with the color gradient indicating temporal progression.
Triangle markers denote the moments when the player fired their weapon, while cross markers represent confirmed eliminations.
\autoref{fig:2d_vis_trajectory} shows the corresponding 2D crosshair trajectory, capturing the physical input behavior mapped to screen space.
To reduce visual clutter from overlapping lines, the temporal color gradient is omitted, while all other markers retain their meanings from \autoref{fig:3d_vis_trajectory}.

\begin{figure*}[htbp]
\centering
\includegraphics[width=1.0\textwidth]{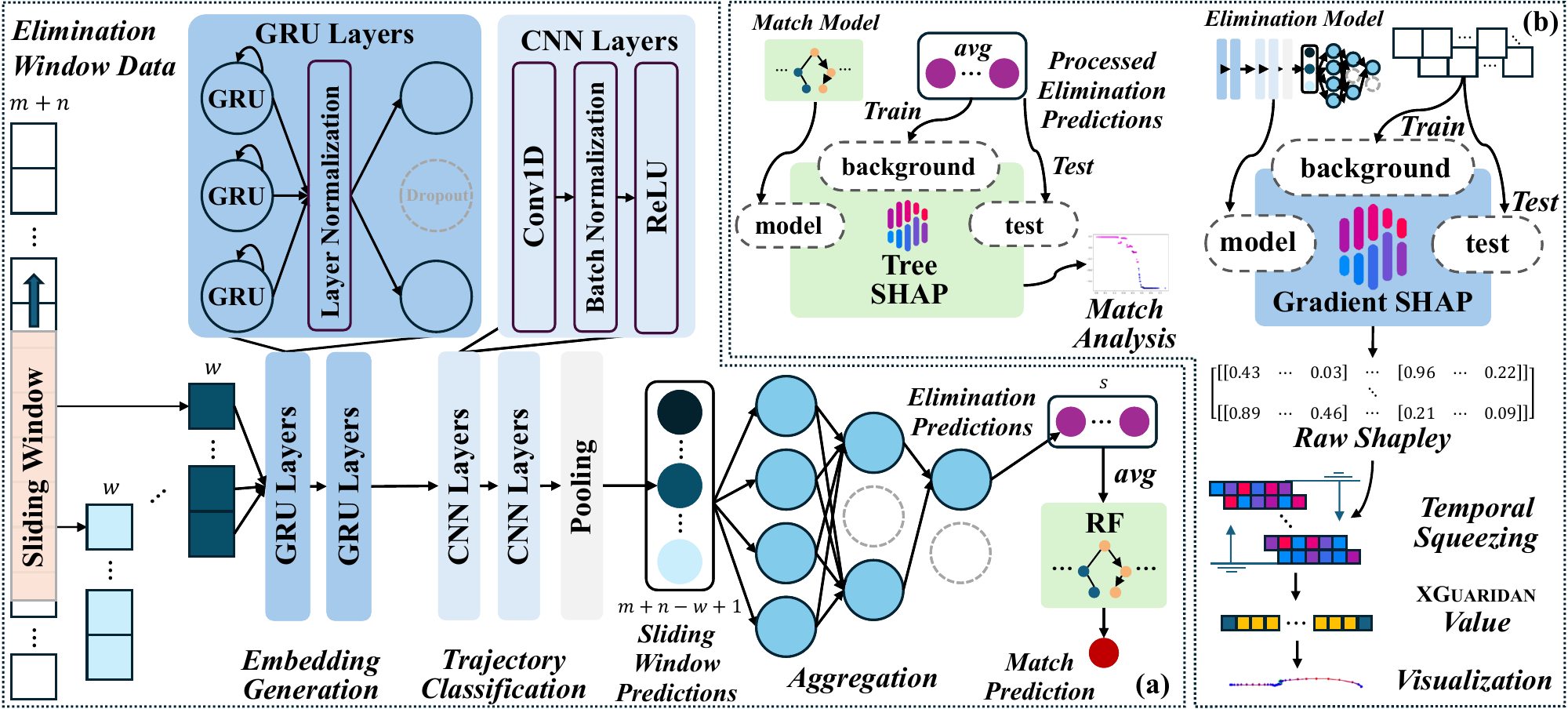}
\caption{\sys framework. The subfigures illustrate the architectures of (a)-\inspector and (b)-\explainer.}
\label{fig:framework}
\end{figure*} 
\subsubsection{Feature construction}
\label{sec:feature_construction}
After obtaining the 2D coordinates, we encounter a challenge: despite aligning the coordinates based on each elimination event, the aiming trajectories remain highly scattered and patternless. 
As a result, directly using the raw trajectory coordinates as input is impractical.
To address this, we design features illustrated in \autoref{fig:2d_feature} that effectively describe a player's aiming trajectory. 
These features must be uncorrelated with the actual spatial distribution of coordinates while providing a standardized, quantifiable representation of the information contained in the aiming trajectory.
These features include the instantaneous velocity and acceleration of trajectory points in the x- and y-directions, as well as the angular change in velocity direction.

\textbf{Velocity} ($\vec{v}$) is calculated as the rate of change of position with respect to time. 
Its components in the x and y directions are computed using the differences in the x and y coordinates divided by the differences in the \textit{tick} values shown in \autoref{eq:vx} and \autoref{eq:vy}.
\begin{equation}\label{eq:vx}
\vec{v}_x = \frac{dx}{dt}
\end{equation}
\vspace{-1em}
\begin{equation}\label{eq:vy}
\vec{v}_y = \frac{dy}{dt}
\end{equation}

\textbf{Acceleration} ($\vec{a}$) is calculated as the rate of change of velocity with respect to time. 
The acceleration components in the x and y directions are computed using the differences in the velocity components divided by the differences in the \textit{tick} values shown in \autoref{eq:ax} and \autoref{eq:ay}.
\begin{equation}\label{eq:ax}
\vec{a}_x = \frac{d\vec{v}_x}{dt} = \frac{d^2x}{dt^2}
\end{equation}
\vspace{-1em}
\begin{equation}\label{eq:ay}
\vec{a}_y = \frac{d\vec{v}_y}{dt} = \frac{d^2y}{dt^2}
\end{equation}

\textbf{Angle of the velocity} ($\alpha$) is calculated using the 2-argument arctangent, which calculates the angle between the positive x-axis and the point given by the coordinates $(\vec{v}_x,\vec{v}_y)$ shown in \autoref{eq:angle}. 
\textbf{Angular change} ($\theta$) is calculated as the difference between the above consecutive angle values. 
This represents the change in the direction of the velocity vector over time shown in \autoref{eq:angle_change}.
\begin{equation}\label{eq:angle}
\alpha = \arctan2(\vec{v}_y, \vec{v}_x)
\end{equation}
\vspace{-1em}
\begin{equation}\label{eq:angle_change}
\theta = \frac{d\alpha}{dt}
\end{equation}

\autoref{fig:2d_feature} illustrates an exemplary case.
The two diamond-shaped points represent the trajectory positions at \textit{ticks},  $t_0$ and $t_1$. 
The solid red line represents the segment of the aiming trajectory between these two points, while the dashed red line represents the trajectory before and after these points.
At \textit{tick} $t_0$, the instantaneous velocity is $\vec{v}_0$, with its direction determined by the line connecting the previous trajectory point to $t_0$. 
The velocity components in the x- and y-directions are then computed, allowing us to further calculate the angle. 
The same process applies to $\vec{v}_1$ at $t_1$.
Using the instantaneous velocities at $t_0$ and $t_1$, we can compute the angular change $\theta_1$ between them. 
Furthermore, we can calculate the instantaneous acceleration and its corresponding components in the x- and y-directions.
It is important to note that since computing the above features requires information from the previous time step, we always discard the first data tuple when calculating features, as its values are not applicable.

\subsubsection{ML-ready data}
\label{sec:ml_ready_data}
After the aforementioned preprocessing, the dataset constructed and utilized in this paper follows the tuple structure $[t,I_{f},I_{e},\vec{v}_x,\vec{v}_y,\vec{a}_x,\vec{a}_y,\theta]$, where $t$ represents the absolute \textit{tick} of the observation, $I_{f}$ and $I_{e}$ are Boolean values indicating whether the player is firing ($I_{f}$) or conducting an elimination ($I_{e}$).
%
The remaining values correspond to the features described in the previous section.
Each valid tuple will be combined into an array to become a sample (i.e., an elimination window) and used as input to \sys.

\subsection{\inspector}
\label{sec:inspector}
\inspector's architecture is shown in \autoref{fig:framework}\textcolor{green!80!black}{a}.
\inspector processes samples composed of $m+n$ ML-ready tuples, as described in \autoref{sec:extraction_scope} and \autoref{sec:ml_ready_data}.
We empirically set the size of each elimination window to 96, with the parameter selection process detailed in \autoref{sec:eval_inspector_para_selection}.
A sliding window mechanism is employed to segment each input sample into $m+n-w+1$ sub-sequences, where $w$ denotes the size of the sliding window.
Based on empirical evaluation, we set $w = 6$, and the selection rationale is also discussed in \autoref{sec:eval_inspector_para_selection}.

GRU is a well-established recurrent architecture known for its training efficiency and competitive performance relative to LSTM \cite{LSTM1997Hochreiter}, while CNNs are effective for pattern recognition and classification tasks.
Each sub-sequence is processed by a hybrid architecture consisting of stacked GRU and CNN layers, as illustrated in \autoref{fig:framework}, which jointly perform temporal embedding and local pattern classification.
This GRU-CNN design demonstrates superior performance over alternative baselines, as presented in \autoref{sec:eval_inspector_para_selection}.
For each elimination window, the model produces $m+n-w+1$ predictions—one for each sub-sequence generated by the sliding window.
These predictions are then embedded and fed into stacked fully connected layers, with dropout applied to reduce overfitting.
Assume a player has $s$ elimination windows within a match.
The $s$ sets of predictions are averaged and passed into a random forest classifier.
We adopt a random forest due to its flexible decision boundaries, which are well-suited to handling the nonlinear structure of the aggregated prediction embeddings, enabling robust and dynamic classification.
Its final binary output, indicating whether the player is cheating, constitutes the decision produced by \sys.

\subsection{\explainer}
\label{sec:explainer}
We choose SHAP over other candidates below for \explainer considering its (a) loose coupling and (b) explanation stability.
Attention weight is model-specific and often correlates loosely with causal feature importance \cite{serrano2019attentioninterpretable}. 
LIME \cite{lime} can be unstable and inconsistent due to random sampling, especially in time-series data with nonlinear dependencies and strong feature interactions (e.g., FPS games).
Instead, SHAP’s additive decomposition compares the influence of temporal features across model variants in unified frameworks. 
Its model-agnostic nature provides loose coupling with the detection framework and thereby makes it easy to evaluate or replace different detection architectures.
\explainer's architecture is shown in \autoref{fig:framework}\textcolor{green!80!black}{b}.
\explainer consists of two main components: the \textit{Elimination Trajectory Explainer} (\autoref{fig:framework}\textcolor{green!80!black}{b} left) and the \textit{Match Explainer} (\autoref{fig:framework}\textcolor{green!80!black}{b} right).
The \textit{Elimination Trajectory Explainer} analyzes the feature importance contribution for each \textit{tick} within an input elimination window.
Meanwhile, the \textit{Match Explainer} provides an interpretation of the decision regarding whether a player engaged in cheating during a match.

Both explainers leverage \textit{GradientExplainer} or \textit{TreeExplainer} from SHAP, as introduced in \autoref{sec:xai}.
%
These explainers are initialized using the trained model and the training dataset (serving as the background for baseline comparison). We then apply the explainers to instances from the test dataset to compute their corresponding Shapley values.
However, since SHAP is primarily designed for structured data, it cannot be directly applied to time-series trajectory data, particularly in the context of the \textit{Elimination Trajectory Explainer}.
To address this, we propose an interpretability enhancement method specifically tailored for temporal data.
For each sample, every feature at each time step within a sliding window receives corresponding Shapley values.
Time steps at the boundaries of an elimination window (i.e., the first and last) yield only one set of Shapley values, while intermediate time steps, due to overlapping sliding windows, may produce between two and six sets.
After applying \textit{GradientExplainer}, the resulting Shapley values for each elimination window form a matrix of shape $(m+n, f)$, where $f$ is the number of features.
Since each sliding window is treated with equal importance in explaining elimination trajectories in an anti-cheat scenario, we introduce a mechanism called \textit{temporal squeezing}.
\textit{Temporal squeezing} aggregates the Shapley values for each feature across all overlapping time steps in the elimination window and computes their mean, producing a set of values referred to as \sys \textit{values}.
These values quantify feature importance at each time step and address the challenge of applying SHAP to temporal data in FPS anti-cheat applications.
\autoref{eq:sys_value} formalizes the computation of the \sys \textit{value} under various conditions within an elimination window, where $SP(f)$ denotes the Shapley value of feature $f$ at a given time step $i$.
The \sys \textit{value} represents the average marginal contribution of a feature toward the model’s prediction.
A higher \sys \textit{value} indicates a greater contribution to a positive prediction (i.e., cheating), whereas a lower or negative value implies influence toward a non-cheating prediction.
For the \textit{Match Explainer}, we adopt a simpler approach by directly using SHAP’s built-in Shapley values for match-level interpretation and visualization.
\autoref{sec:case_study} presents and visualizes both the trajectory-wise and match-wise explanation results.
\vspace{-0.5em}
\begin{equation}\label{eq:sys_value}
    \text{\textbf{V}$_\sys$}(f, i) =
    \begin{cases}
        SP(f), & i = 1 \text{ or } i = m+n \\
        \frac{\sum SP(f)}{i}, & 2 \leq i \leq w \\
        \frac{\sum SP(f)}{w}, & w+1 \leq i \leq m+n-w \\
        \frac{\sum SP(f)}{m+n-i}, & m+n-w+1 \leq i \leq m+n-1
    \end{cases}
\end{equation}
\section{Experiments}
\label{sec:evaluation}
\vspace{0.3em}
\subsection{Experimental Setups}
\label{sec:experimental_setups}

\subsubsection{Dataset Construction}
\label{sec:dataset_construction}
\noindent\textbf{Data Collection.}
This study is conducted in collaboration with \textit{5EPlay\footnote{5EPlay: \href{https://arena.5eplay.com/}{https://arena.5eplay.com/}.}}, a world-class CS2 gaming platform. 
Players engage in regular gameplay on the platform, and after each match, \textit{demos} are automatically generated. 
The proprietary anti-cheat system is used only for preliminary filtering and does not directly determine the final labels.
Each player is labeled to indicate their cheating status per match.
All labels are manually re-verified before being shared with us.
Consistent with industry standards, labels are binary (i.e., cheater/normal player) and do not distinguish between specific cheat subtypes.
We released all raw \textit{demos} used in this study to ensure long-term reproducibility.
A carefully curated dataset, representing the most accurately labeled data currently achievable in the industry, is then used for our evaluations.

\noindent\textbf{Demo Parsing.}
We use an open-source library, \textit{awpy2} \cite{awpy}, which extracts raw time-series data from the provided CS2 \textit{demos} and parses them into tabular data. 

\noindent\textbf{Trajectory Extraction.}
In terms of CS2, we use its default resolution 1920 $\times$ 1080 as \autoref{alg:py2xy}'s inputs.
Noted that the default resolution is used for normalizing and visualizing the trajectory. 
Different resolutions on the client side would not affect the server-side trajectory conversion, since we are using the server-side pitch-yaw for the reconstruction.

\noindent\textbf{Data Cleansing.} 
When constructing the dataset as described in \autoref{sec:feature_construction}, we filter out certain edge cases to minimize noise.
First, data from players who have no eliminations are removed, as it is impossible to extract elimination windows from such cases.
Second, elimination windows with a length shorter than $m+n$ are discarded. 
This rare condition arises when, within $n$ ticks after an elimination, one of the following occurs: (1) the player disconnects from the server, (2) the match ends immediately, and (3) the match enters a side-switch phase.
Additionally, within the elimination window, we filter out cases where: (4) a server glitch led to missing or duplicated tuples.
The first three cases prevent the player from moving their avatar's view within $n$ ticks after the elimination, while the last one disrupts the temporal sequence of the data. 
Both could contaminate the dataset and thereby be discarded.

\noindent\textbf{Dataset Description.} 
\autoref{tab:dataset_description} provides a description of the dataset we constructed, including three data splits.
When we obtain the \textit{demos}, we randomly divide all demos into three equal parts and apply the aforementioned preprocessing steps. 
This process results in the three data splits, which are used as training, validation, and test sets in the following evaluations.
We publicize our repository and dataset. To our knowledge, this is the first publicly available large-scale aiming trajectory dataset for aim-assist cheats detection obtained from a real-world active modern FPS game.
\begin{table}[htbp]
\centering
\caption{Dataset description.}
\vspace{-0.5em}
\label{tab:dataset_description}
\resizebox{\columnwidth}{!}{%
\begin{tabular}{@{}l lll@{}}
\toprule
  \multicolumn{1}{c}{\textbf{Data Split}} & \multicolumn{1}{c}{\textbf{A}} & \multicolumn{1}{c}{\textbf{B}} & \multicolumn{1}{c}{\textbf{C}}      \\ \midrule
\textbf{\#Tick}                & 992,064  & 1,014,624  & 1,062,528 \\
\textbf{\#Total Elimination Window} & 10,334   & 10,569     & 11,068    \\
\textbf{\#Cheating Elimination Window} & 2,385    & 2,760      & 2,900     \\
\textbf{\#Normal Elimination Window} & 7,949    & 7,809      & 8,168     \\
\textbf{\#Match}               & 971      & 944        & 988       \\
\textbf{\#Cheater}             & 108      & 115        & 127       \\
\textbf{\#Normal Player}       & 1,711    & 1,671      & 1,754     \\
\textbf{\#Total Player}        & 1,819    & 1,786      & 1,881     \\ \bottomrule
\end{tabular}%
}
\end{table}



\subsubsection{Training Configurations}
\label{sec:training_config}
\inspector's hyperparameters and configurations for the models are listed in \autoref{appx:model_param}.
The GRU-CNN model is trained with parallelized computation using the mirrored strategy.
Three callbacks are employed during training to improve performance and generalization:
(1) \textit{ModelCheckpoint}, which saves the model with the lowest validation loss,
(2) \textit{ReduceLROnPlateau}, which reduces the learning rate by half if the validation loss plateaus (with a patience of 20 epochs and a minimum learning rate of 0.0001), and
(3) \textit{EarlyStopping}, which halts training if no improvement is observed after 50 consecutive epochs to prevent overfitting.
The aggregation model also utilizes the same \textit{EarlyStopping} mechanism for consistency.
Both models are optimized using the Adam optimizer and trained with a \textit{binary crossentropy} loss function.
We use a batch size of 256 and train for up to 500 epochs, applying class weight balancing to address the class imbalance inherent in the cheating detection task.
The training and validation loss curves are provided in \autoref{appx:loss}.

\subsection{Detection Performace}
\label{sec:detection_performance}
Can the \inspector in \sys effectively identify cheaters? 
To evaluate \sys's anti-cheat performance in a real-world setting, we conduct experiments using the collected dataset.
We apply 3-fold cross-validation to assess \sys's effectiveness and robustness. 
\autoref{tab:overall_performance} presents the overall anti-cheat performance of \sys. 
The three data splits are used as the training, validation, and test sets in different orders to eliminate any potential bias that might arise from specific dataset compositions.
Since the dataset has an imbalanced distribution of cheaters and legitimate players, all evaluation metrics are computed using weighted averages to more accurately reflect the system's real-world performance.
A detailed introduction to the metrics is presented in \autoref{appx:metrics}.
In \autoref{tab:overall_performance}, we observe that \sys achieves high average performance across all metrics, with low variance and standard deviation, indicating that it maintains stable performance across different data splits.
In the following evaluations, we use the data split combination of ABC as default for the training, validation, and test sets, respectively.

\definecolor{train}{HTML}{0F6292}
\definecolor{val}{HTML}{15B392}
\definecolor{test}{HTML}{A0153E}

\begin{table}[htbp]
\centering
\caption{\sys weighted average overall performance with different set combinations. \textbf{\textcolor{train}{Blue}}, \textbf{\textcolor{val}{green}}, and \textbf{\textcolor{test}{red}} in the combination rows indicate the use of the set as training, validation, and test set, respectively.}
\vspace{-0.5em}
\label{tab:overall_performance}
\resizebox{\columnwidth}{!}{%
\begin{tabular}{@{}ccccccc@{}}
\toprule
\multicolumn{1}{l}{} & \multicolumn{1}{l}{} & \textbf{Accuracy$(\uparrow)$} & \textbf{FPR$(\downarrow)$} & \textbf{Recall$(\uparrow)$} & \textbf{Precision$(\uparrow)$} & \textbf{F1-Score$(\uparrow)$} \\ \midrule
\multicolumn{1}{c|}{\multirow{6}{*}{\rotatebox{90}{\textbf{Combination}}}} & \textbf{\textcolor{train}{A}\textcolor{val}{B}\textcolor{test}{C}} & 0.889 & 0.062 & 0.889 & 0.938 & 0.907 \\
\multicolumn{1}{c|}{} & \textbf{\textcolor{train}{A}\textcolor{val}{C}\textcolor{test}{B}} & 0.876 & 0.049 & 0.876 & 0.951 & 0.901 \\
\multicolumn{1}{c|}{} & \textbf{\textcolor{train}{B}\textcolor{val}{A}\textcolor{test}{C}} & 0.882 & 0.101 & 0.882 & 0.899 & 0.890 \\
\multicolumn{1}{c|}{} & \textbf{\textcolor{train}{B}\textcolor{val}{C}\textcolor{test}{A}} & 0.907 & 0.041 & 0.907 & 0.959 & 0.924 \\
\multicolumn{1}{c|}{} & \textbf{\textcolor{train}{C}\textcolor{val}{A}\textcolor{test}{B}} & 0.872 & 0.050 & 0.872 & 0.950 & 0.898 \\
\multicolumn{1}{c|}{} & \textbf{\textcolor{train}{C}\textcolor{val}{B}\textcolor{test}{A}} & 0.879 & 0.048 & 0.879 & 0.952 & 0.904 \\
\cdashline{0-6}[1pt/1pt]
\multicolumn{2}{c}{\textbf{mean} ($\mu$)} & 0.884 & 0.059 & 0.884 & 0.942 & 0.904 \\
\multicolumn{2}{c}{\textbf{var} ($\sigma^{2}$)} & 1.32E-04 & 4.00E-04 & 1.32E-04 & 4.00E-04 & 1.08E-04 \\
\multicolumn{2}{c}{\textbf{std} ($\sigma$)} & 1.15E-02 & 2.00E-02 & 1.15E-02 & 2.00E-02 & 1.04E-02 \\ \bottomrule
\end{tabular}%
}
\end{table}

\subsection{Feature Ablation Study}
\label{sec:feature_ablation_study}
Can the features used in \sys be further simplified or improved?
As discussed in \autoref{sec:ml_ready_data}, \sys relies on a set of input features. 
Among these features, $I_{f}$ and $I_{e}$ represent whether the player fired or achieved an elimination at a given tick. 
Both features are critical to aiming trajectory to representation and therefore cannot be further modified or removed.
To verify the remaining features constitute the optimal configuration, we conduct ablation experiments (\autoref{tab:feature_ablation}) and comparative experiments (\autoref{tab:tick_comp}). 
%

\autoref{tab:feature_ablation} presents the results of the feature ablation study of \sys.
\sys demonstrates consistently superior performance across all metrics compared to individual feature combinations (first three rows).
The results of pairwise feature combinations (next three rows) indicate that no single combination achieves optimal performance across all metrics, and this holds when compared to \sys as well. 
For example, while the combination of velocities and angle change yields the highest accuracy, its FPR and precision are worse than those of the other two combinations.
Therefore, to comprehensively evaluate the performance of \sys against pairwise combinations, we need to compare the average performance across all pairwise combinations with \sys.
By averaging the results of the pairwise combinations, \sys outperforms the average performance of all pairwise combinations across all metrics. 
Consequently, we choose to utilize the full set of features.

\autoref{tab:tick_comp} presents the comparative results of different modifications or removals of the tick feature. 
Specifically, the raw tick index refers to the original tick values from the dataset (e.g., within a given elimination window, raw tick indices might range from 5000 to 5095); the reset tick index replaces the raw tick index with a fixed range of 0 to 95.
%
%
Evaluation results show that using the raw tick index outperforms the alternative approaches. 
This is because some cheaters exhibit delayed activation behaviors, where cheats are enabled only after falling behind during a match. 
Using raw ticks preserves such mid-match temporal context, allowing the model to better capture phase-dependent behavioral shifts.

\begin{table}[htbp]
\centering
\caption{Feature ablation study weighted average results.}
\vspace{-0.5em}
\label{tab:feature_ablation}
\resizebox{\columnwidth}{!}{%
\begin{tabular}{@{}ccc|ccccc@{}}
\toprule
\multicolumn{3}{c|}{\textbf{Feature}} & \multicolumn{5}{c}{\textbf{Metrics}} \\ 
\midrule
\textbf{$\vec{a}_x$,$\vec{a}_y$} & \textbf{$\vec{v}_x$,$\vec{v}_y$} & \textbf{$\theta$} & \textbf{Accuracy} & \textbf{FPR} & \textbf{Recall} & \textbf{Precision} & \textbf{F1-Score} \\ 
\midrule
$\checkmark$ & $\usym{2613}$ & $\usym{2613}$ & 0.838 & 0.059 & 0.838 & 0.941 & 0.873 \\
$\usym{2613}$ & $\checkmark$ & $\usym{2613}$ & 0.884 & 0.060 & 0.884 & 0.940 & 0.904 \\
$\usym{2613}$ & $\usym{2613}$ & $\checkmark$ & 0.877 & 0.057 & 0.877 & 0.943 & 0.900 \\
\cdashline{0-7}[1pt/1pt]
$\checkmark$ & $\checkmark$ & $\usym{2613}$ & 0.886 & 0.062 & 0.886 & 0.938 & 0.905 \\
$\checkmark$ & $\usym{2613}$ & $\checkmark$ & 0.880 & 0.058 & 0.880 & 0.942 & 0.902 \\
$\usym{2613}$ & $\checkmark$ & $\checkmark$ & 0.901 & 0.063 & 0.901 & 0.937 & 0.915 \\
\cdashline{0-7}[1pt/1pt]
$\checkmark$ & $\checkmark$ & $\checkmark$ & 0.890 & 0.060 & 0.890 & 0.940 & 0.908 \\ \bottomrule
\end{tabular}%
}
\end{table}
\begin{table}[htbp]
\centering
\caption{\sys input feature \textit{tick} indexing comparative analysis weighted average results. `w/o' and `w/' denote without and with. `Reset index' indicates re-allocating each elimination window's tick index from 0 to $m+n-1$.}
\vspace{-0.3em}
\label{tab:tick_comp}
\resizebox{\columnwidth}{!}{%
\begin{tabular}{@{}lccccc@{}}
\toprule
\textbf{Tick Indexing}                                & \textbf{Accuracy} & \textbf{FPR}   & \textbf{Recall} & \textbf{Precision} & \textbf{F1-Score} \\ \midrule
w/o tick index                               & 0.801             & 0.078          & 0.801           & 0.922              & 0.846             \\
w/ reset index                               & 0.784             & 0.073          & 0.784           & 0.927              & 0.834             \\
\textbf{\sys (w/ raw index)} & \textbf{0.889}    & \textbf{0.062} & \textbf{0.889}  & \textbf{0.938}     & \textbf{0.907}    \\ \bottomrule
\end{tabular}%
}
\end{table}

\subsection{\inspector Model and Parameters}
\label{sec:eval_inspector_para_selection}
Are the models and parameters in \sys optimal for the aim-assist cheats detection?
To justify our choice, we evaluated the performance of \sys using different models and parameters.
Recapture in \autoref{sec:extraction_scope}, \autoref{sec:inspector}, \autoref{fig:elimination_window}, and \autoref{fig:framework}, \sys selects an elimination window of size $m+n$, capturing $m$ ticks before and $n$ ticks after the elimination tick. 
We investigate the impact of the elimination window size on \sys's performance in \autoref{sec:diff_elimination_win_size}.
\sys employs a length of $w$ sliding window for segmenting elimination information.
We discover the different performance of \sys with different lengths of the sliding window in \autoref{sec:diff_sliding_win_size}.
\sys leverages the GRU-CNN model for the embedding generation and trajectory classification.
We evaluate the different performance of \sys with different models in \autoref{sec:diff_model_embedding_gen} and \autoref{sec:diff_model_trajectory_cls}.
\sys integrates the elimination predictions through the averaging operation to integrate and generate a match prediction.
We investigate the influence of using different representations of elimination predictions embedding on the performance of \sys in \autoref{sec:diff_style_elimination_pred}.

\subsubsection{Different Elimination Window Sizes}
\label{sec:diff_elimination_win_size}
%
In \autoref{fig:elimination_window_selection}, variations in the elimination window size have a minimal effect on cheat detection performance (a maximum 0.4\% difference between the optimal and the chosen value). 
Considering that larger windows exponentially increase system overhead (including storage, computation, memory, and processing time for preprocessing, training, and inference), we choose a window size of 96 as \sys's parameter.
Since CS2’s average time-to-kill is approximately 0.5 seconds, a 96-tick ($\approx$ 1.5 seconds) window ensures full coverage of pre-aiming, firing, and elimination behaviors. Smaller windows would truncate critical firing events and thus lead to the loss of essential trajectory information.

\subsubsection{Different Sliding Window Sizes}
\label{sec:diff_sliding_win_size}
\begin{figure}[htbp]
\centering
\begin{subfigure}[htbp]{0.23\textwidth}
    \centering
    \includegraphics[width=\textwidth]{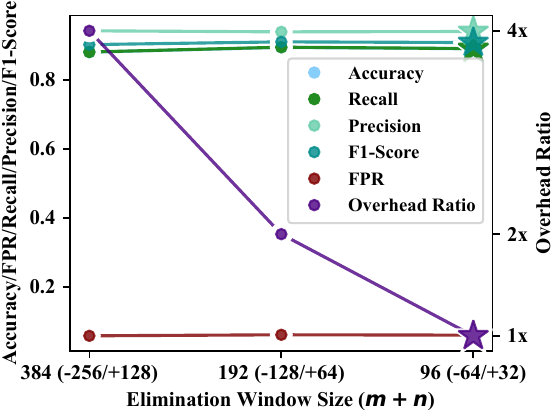}
    \vspace{-0.5em}
    \caption{Elimination window}
    \label{fig:elimination_window_selection}
\end{subfigure}
\hfill
\begin{subfigure}[htbp]{0.24\textwidth}
    \centering
    \includegraphics[width=\textwidth]{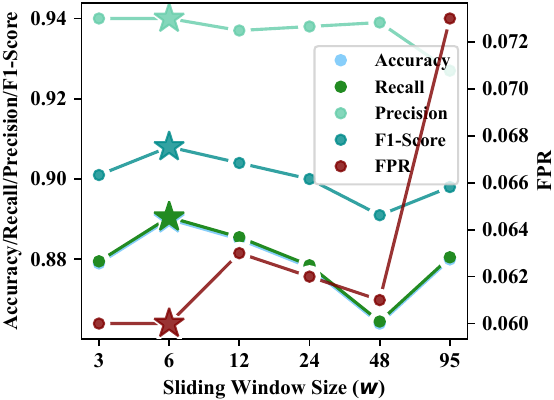}
    \vspace{-0.5em}
    \caption{Sliding window}
    \label{fig:sliding_window_selection}
\end{subfigure}
\caption{\inspector sliding window and elimination window size comparative analysis weighted average results. The star ($\filledstar$) indicates the parameters used by \sys.}
\label{fig:window_selections}
\end{figure}

In \autoref{fig:sliding_window_selection}, for different evaluation metrics, we empirically obtain that the optimal sliding window size is 6. 
%

\begin{figure}[htbp]
\centering
\begin{subfigure}[htbp]{0.235\textwidth}
    \centering
    \includegraphics[width=\textwidth]{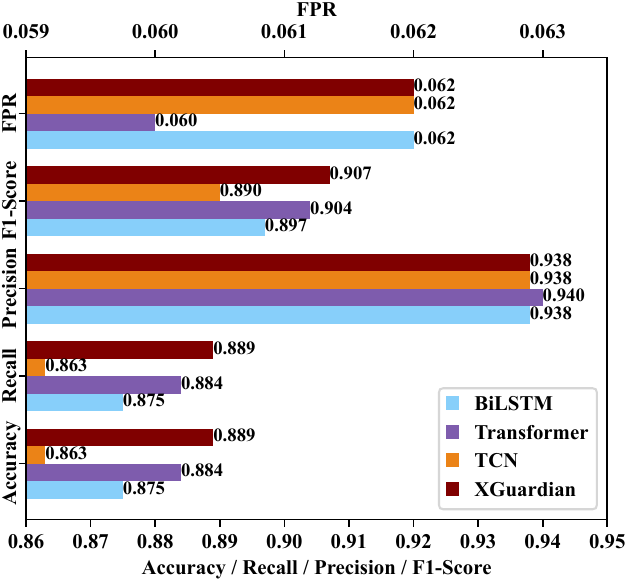}
    \vspace{-0.5em}
    \caption{Embedding generation stage}
    \label{fig:embedding_generation_baseline_comparison}
\end{subfigure}
\hfill
\begin{subfigure}[htbp]{0.235\textwidth}
    \centering
    \includegraphics[width=\textwidth]{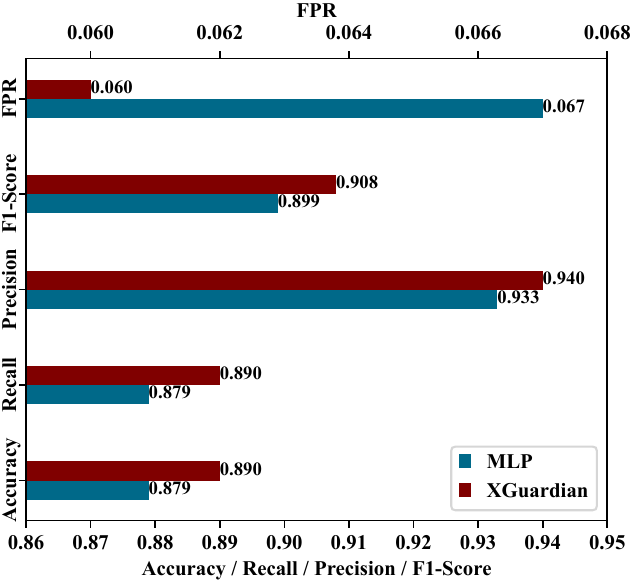}
    \vspace{-0.5em}
    \caption{Trajectory classification stage}
    \label{fig:trajectory_classification_baseline_comparison}
\end{subfigure}
\caption{\inspector embedding generation and trajectory classification stages model selection weighted average results across SOTA models \cite{tcn,bilstm,transformer,MLPs}.}
\label{fig:inspector_model_selection}
\vspace{-0.5em}
\end{figure}

\subsubsection{Different Models in Embedding Generation}
\label{sec:diff_model_embedding_gen}
As shown in \autoref{fig:embedding_generation_baseline_comparison}, we compare various state-of-the-art time-series models, BiLSTM \cite{bilstm}, Transformer \cite{transformer}, and TCN \cite{tcn}. 
While the Transformer achieves a slightly higher precision than the GRU used in \sys (by 0.2\%), \sys outperforms all other models across the remaining metrics.
Additionally, the computational overhead of the Transformer is significantly higher than that of GRU \cite{gru_is_faster_than_transformer}. 
Therefore, we retain the current GRU-based structure in the embedding generation stage of \sys.

\subsubsection{Different Models in Trajectory Classification}
\label{sec:diff_model_trajectory_cls}
As shown in \autoref{fig:trajectory_classification_baseline_comparison}, we compare the CNN used in \sys with MLPs \cite{MLPs} (i.e., fully connected layers). 
The result demonstrates that \sys achieves the best performance across all evaluation metrics.

\subsubsection{Different Input Styles in Elimination Predictions}
\label{sec:diff_style_elimination_pred}
In \autoref{tab:elimination_predictions_embedding_style_comp}, we compare all possible representations of elimination predictions. 
Among these, \textit{binary} refers to the binary classification of elimination predictions. 
Specifically, the binary classification is determined by a dynamic threshold, the value of which is learned from the training and validation sets. 
The results show that using the original predicted values as input to the match prediction model yields better overall performance than the other combinations.

\begin{table}[htbp]
\centering
\caption{\inspector elimination predictions stage embedding style comparison. `Binary' denotes embedding binary classified predictions as match prediction's input. `Original' denotes embedding raw predictions as match prediction input. \sys's configuration is bolded.}
\vspace{-0.5em}
\label{tab:elimination_predictions_embedding_style_comp}
\resizebox{\columnwidth}{!}{%
\begin{tabular}{@{}lccccc@{}}
\toprule
\textbf{Input Style} & \textbf{Accuracy} & \textbf{FPR} & \textbf{Recall} & \textbf{Precision} & \textbf{F1-Score} \\ \midrule
Binary               & 0.881             & 0.061        & 0.881           & 0.939              & 0.902             \\
Binary+Original      & 0.881             & 0.061        & 0.881           & 0.939              & 0.902             \\
\textbf{\sys (Original)}    & \textbf{0.889}    & \textbf{0.062} & \textbf{0.889}  & \textbf{0.938}     & \textbf{0.907}    \\ \bottomrule
\end{tabular}%
}
\end{table}

\subsection{Comparative Study}
\label{sec:compare_sota}
Is \sys performing better compared to prior works?
We compare \sys with seven other server-side detection methods, including comparisons with two other reproduced methods using the dataset utilized by \sys. 
Due to the lack of available open-source solutions, we reproduced two server-side aim-assist cheat detection methods based on simple statistical features related to the aiming or elimination used in prior works. 
Both methods are related to \sys but use non-temporal features, data, and approach. Below is a summary of each method:

\begin{itemize}[topsep=2pt, itemsep=2pt, parsep=0pt, partopsep=0pt]
\item \textsc{th\_HitAcc}: Filters cheaters based on the weapon accuracy of the player, using a threshold \cite{alayed2013ml,galli2011unreal}.
\item \textsc{th\_AccA}: Filters cheaters when the player's field-of-view acceleration exceeds a threshold \cite{yu2012statistical}.
\end{itemize}

Furthermore, we compare \sys with the state-of-the-art server-side solution, \textsc{Hawk}, and its three subsystems using the aimbot dataset provided by the same solution \cite{hawk2024}. Below is a summary of the method and its sub-systems:
\begin{itemize}[topsep=2pt, itemsep=2pt, parsep=0pt, partopsep=0pt]
\item \textsc{RevPOV}: Uses LSTM to classify a wide range of temporal operation data of the player for detecting cheaters \cite{hawk2024}.
\item \textsc{RevStats}: Designs structured features to represent the player's behavior and classify with ensemble learning \cite{hawk2024}.
\item \textsc{ExSPC}: Uses deep learning to examine the consistency of a player's gaming sense and their combat performance \cite{hawk2024}.
\item \textsc{Hawk}: Combines the judgments from the above three subsystems and dynamically classifies cheaters using a threshold that can quickly adapt to anti-cheat requirements \cite{hawk2024}.
\end{itemize}

The results are shown in \autoref{tab:sota_comp}.
\sys outperforms previous works on both datasets. 
%
%
The reproduction results of the prior works are aligned with their paper performance (if applicable on the large dataset).
A direct comparison of \textsc{Hawk} with \sys's dataset is infeasible, as \textsc{Hawk}'s features are highly game-specific, as discussed in \autoref{sec:introduction}, and some of the raw data (e.g., whether the opponents are in sight) required for feature extraction are not available in CS2's \textit{demo}.

\begin{table}[htbp]
\centering
\caption{\sys comparative study to prior works using different data sources. The optimal results are bolded.}
\vspace{-0.5em}
\label{tab:sota_comp}
\resizebox{\columnwidth}{!}{%
\begin{tabular}{@{}llccccc@{}}
\toprule
\textbf{Dataset}               & \textbf{Method}                          & \textbf{Accuracy} & \textbf{FPR}   & \textbf{Recall} & \textbf{Precision} & \textbf{F1-Score} \\ \midrule
\multirow{3}{*}{\textbf{\sys}} 
    & \textsc{th\_AccA} \cite{yu2012statistical}  & 0.738             & 0.455          & 0.738           & 0.545              & 0.627             \\
    & \textsc{th\_HitAcc} \cite{galli2011unreal,alayed2013ml} & 0.644             & 0.086          & 0.644           & 0.914              & 0.711             \\
    & \textbf{\sys}                                & \textbf{0.889}    & \textbf{0.062} & \textbf{0.889}  & \textbf{0.938}     & \textbf{0.907}    \\ \midrule
\multirow{5}{*}{\textbf{\textsc{Hawk}} \cite{hawk2024}}      
    & \textsc{RevPOV} \cite{hawk2024}       & 0.616             & 0.151          & 0.616           & 0.849              & 0.694             \\
    & \textsc{RevStats} \cite{hawk2024}     & 0.814             & 0.109          & 0.814           & 0.891              & 0.842             \\
    & \textsc{ExSPC} \cite{hawk2024}        & 0.793             & 0.114          & 0.793           & 0.886              & 0.827             \\
    & \textsc{Hawk} \cite{hawk2024}              & 0.716             & 0.112          & 0.716           & 0.888              & 0.772             \\
    & \textbf{\sys}                                & \textbf{0.841}         & \textbf{0.056}      & \textbf{0.841}       & \textbf{0.944}          & \textbf{0.878}         \\ \bottomrule
\end{tabular}%
}
\end{table}

\subsection{Generalizability Study}
\label{sec:generalizability}
Can \sys be generalized to different games developed with different game engines? 
\autoref{tab:generalizability} demonstrates \sys's generalizability evaluation results.
We observe that \sys's performance remains consistent on different subgenres of games, engines, and platforms. 
Note that even though CS:GO and CS2 are from the same series, they use entirely different engines with different replay systems and parsers, and thereby should be treated as two different games.
We note a slight decrease in the effectiveness of \sys on the Farlight84 dataset due to the fact that the precision of its playback system (i.e., the frequency of operations recording) is much lower than that of the CS2's (approximately half that of the CS2's).
Therefore, the trajectory description is vaguer than the CS2 dataset and thereby leads to a decrease.
Farlight84 data is provided by its developer \textit{Lilith Games}\footnote{Lilith Games: \href{https://ancient.lilith.com/en/}{https://ancient.lilith.com/}.}.
\sys's performance slightly decreases on \textsc{Hawk}'s dataset due to the existence of inaccurate negative labels in the dataset, as mentioned by their authors.
Overall, the performance of \sys in this experiment is acceptable in terms of evaluating the generalizability.

\begin{table}[htbp]
\centering
\caption{\sys's generalizability evaluation results on different sub-genre games, platforms, and engines. The descriptions of the other two datasets are shown in \autoref{appx:generalizability_dataset_description}.}
\vspace{-0.5em}
\label{tab:generalizability}
\resizebox{\columnwidth}{!}{%
\begin{tabular}{@{}llll@{}}
\toprule
\textbf{Dataset} & \textbf{\textsc{Hawk}} \cite{hawk2024} & \textbf{\textsc{Farlight84}} & \textbf{\sys} \\ \midrule
\textbf{Game (Release Year)} & CS:GO (2012) & Farlight84 (2023) & CS2 (2023) \\
\textbf{Sub-genre} & Tactical Shooter & Battle Royal & Tactical Shooter \\
\textbf{Engine} & Source & Unreal Engine 4 & Source 2 \\
\textbf{Platform} & PC & Mobile & PC \\
\textbf{Accuracy} & 0.841 & 0.816 & 0.889 \\
\textbf{FPR} & 0.056 & 0.144 & 0.062 \\
\textbf{Recall} & 0.841 & 0.816 & 0.889 \\
\textbf{Precision} & 0.944 & 0.856 & 0.938 \\
\textbf{F1-Score} & 0.878 & 0.829 & 0.907 \\ \bottomrule
\end{tabular}%
}
\end{table}

\subsection{Explainability Case Study}
\label{sec:case_study}
The elimination-scope cases analyze how \sys arrives at individual elimination predictions. 
The match-scope case illustrates how \sys generates a final match-level prediction, using a representative cheater case for demonstration.
%
%
Video demonstrations and interactive analyses for all case studies are available on our project website\footnote{Case study website: \href{https://xguardian-anti-cheat.github.io/}{https://xguardian-anti-cheat.github.io/}.}.
To ensure the selected cases are representative and insightful, they are curated by two FPS game experts whose skill levels are detailed in \autoref{tab:expert_skill_level}.
Case A, C, and E are blatant cheaters; Case D is a disguised cheater; Case B is a normal player.
A detailed discussion on the cases can be found either on the website or in \autoref{appx:case_study_addtional}.
Particularly, in Case D, the cheater only employs wallhack. 
Legitimate players scan and react to visual cues. 
A wallhack user, however, knows an opponent's location in advance. 
\sys detects wallhacks by identifying the anomalous aiming behavior they produce.
This often manifests as low-valued features just before an engagement; the cheater simply waits for the target to cross their pre-positioned crosshair.
\sys detects this unnatural lack of movement, which is counterintuitive compared to high-valued movements while using aimbots, but is an indicative signal of wallhacking.

\begin{table}[htbp]
\centering
\caption{Experts' skill level for case study selection and review, adversarial sample selection, and user study case selection. All stats are retrieved from \cite{tracker}.}
\vspace{-0.5em}
\label{tab:expert_skill_level}
\resizebox{\columnwidth}{!}{%
\begin{tabular}{@{}cll@{}}
\toprule
\multicolumn{1}{l}{\textbf{Expert}} & \multicolumn{1}{l}{\textbf{Playtime}}      & \multicolumn{1}{l}{\textbf{Skill Level}}                                                                                                              \\ \midrule
A              & \textgreater{}7,000hrs & \begin{tabular}[c]{@{}l@{}}CS2 - \textit{Premier} 19,000 (Top 2.62\%)\\ Overwatch2 - \textit{Role} Grand Master (Top 1.46\%)\\ Apex Legend - \textit{BR} Master (Top 0.6\%)\end{tabular} \\
B              & \textgreater{}5,000hrs & CS2 - \textit{Premier} 18,000 (Top 4.59\%) \\ \bottomrule
\end{tabular}%
}
\end{table}

\subsection{Robustness Under Adversarial Attack}
\label{sec:adversarial_attack}
Suppose that cheaters know \sys's mechanism and attempt to evade detection, will \sys be dysfunctional?
Theoretically, cheaters must fundamentally alter their cheating behaviors, such as prolonging aim-locking durations, smoothing out jerky movements, or avoiding unnaturally perfect pre-aiming. 
These lead to two outcomes: (1) If they mimic normal behavior but still desire illegal advantages, as in Case A and D, they will still be detected by \sys; (2) If they perfectly mimic legitimate player behaviors, the cheat itself becomes ineffective, as normal players can outperform `\textit{normal players}' with such limited assistance. 
This defeats the purpose of cheating. 
This is a cornerstone of behavioral-based security and has been noted in prior works \cite{choi2023botscreen, hawk2024}.

To validate \sys's robustness in detecting cheaters who attempt to conceal their behavior by mimicking legitimate players, we first computed the mean values of each feature for all non-cheaters on the test set. 
These averages served as a baseline representation of normal gameplay behavior.
Next, for each cheater per match, we evaluated whether its feature value fell within $\pm10\%$ of the corresponding non-cheating averages. 
The final adversarial dataset contains cheaters per match that meet this similarity condition under one or more features in the test set.
We further discuss the parameter selections in \autoref{appx:adversarial_config}.
We have collected 22 cheaters with 54,816 ticks of data.
We further invite the two game experts to verify and confirm the stealthiness of all selected samples.
\sys identifies 18 among 22 cheaters, achieves 0.818 recall and 0.900 F1-Score.
Both the two‐proportion Z‐test \cite{neyman_pearson_1933} and Fisher’s exact test \cite{fisher_1922_interpretation} indicate that the observed decrease in recall under adversarial attack is statistically insignificant ($p>0.05$).
Note that due to our threat model settings, we are unable to acquire the cheater's specific (a) cheat functions and (b) parameter configurations.
\subsection{Overhead}
\label{sec:overhead}
\sys is designed as an asynchronous system that analyzes post-game \textit{replay files} during periods of low server activity.
It operates independently of active game servers and can be deployed on a separate server.
We conduct overhead assessments for reference.
We utilize an Ubuntu 22.04.5 LTS server equipped with 24-core Intel(R) Xeon(R) Gold 5418Y and 503 GB RAM, a real dedicated game server configuration \cite{forum_serverreq}.
We use two NVIDIA GeForce RTX 4090 (24GB) for the model training and testing.
\sys can train with one GPU with over 12 GB memory, but we adopt the mirror strategy in all experiments for synchronous distributed training.
Leveraging the pre-trained model, \sys requires only CPU resources to function efficiently. 
\autoref{tab:overhead} shows the overhead statistics on the entire dataset.
We further evaluate \sys's server overhead in a real-world deployment scenario. Based on data from our partner platform, the system handles a maximum of 150,000 matches daily, with each individual server processing up to 573 matches per day.
Analysis of platform metrics and public server statistics \cite{steamdb2024} reveals that player activity reaches its lowest point between 1:00 AM and 5:00 AM local time. 
During this off-peak window, the system must process a maximum of 144 matches per hour.
Our performance measurements indicate that \sys can complete all required analyses within this timeframe.
Based on the statistics in \autoref{tab:overhead}, the total prediction time for the entire dataset (2,903 matches across three splits from \autoref{tab:dataset_description}) is 28,966.87 seconds, so it takes only less than 9.98 seconds of prediction time per match with CPU-only.
This represents a remarkable improvement of 4,104.31\% compared to the state-of-the-art server-side solution \textsc{Hawk} \cite{hawk2024}, which requires over 400 seconds. 
Furthermore, \sys demonstrates superior memory efficiency, consuming less than 8.6GB RAM for the entire dataset, whereas \textsc{Hawk} demands over 5GB RAM per single match.
Therefore, each server has more than sufficient resources for deploying \sys.
\begin{table}[htbp]
\centering
\caption{\sys's overheads tested with the entire dataset. The numbers are retrieved by the Python libraries time \cite{python-time} and resource \cite{python-resource}.}
\label{tab:overhead}
\vspace{-0.5em}
\resizebox{\columnwidth}{!}{%
\begin{tabular}{@{}llll@{}}
\toprule
\textbf{Overheads} & \textbf{Preprocessing} & \textbf{Prediction} & \textbf{Training} \\ \midrule
\textbf{User time (sec)} & 52.4 & 28966.87 & 28960.42 \\
\textbf{System time (sec)} & 30.21 & 2905.86 & 2905.14 \\
\textbf{Elapsed time (sec)} & 48.03 & 3.95 & 11423.11 \\
\textbf{Max. resident set size (GB)} & 3.226 & 8.596 & 8.596 \\ \bottomrule
\end{tabular}%
}
\end{table}

\subsection{User Study on \explainer's Bias}
\label{sec:user_study}

\definecolor{lightgrayD}{HTML}{D9D9D9}
\setlength{\fboxsep}{1pt}
\begin{table*}[htbp]
\caption{User study results of the \explainer's potential bias questionnaire.}
\vspace{-0.5em}
\label{tab:user_study_result}
\begin{threeparttable}
\resizebox{\textwidth}{!}{%
\begin{tabular}{@{}ccccccccccccccccccccc@{}}

\toprule
                                    & \multicolumn{6}{c}{\textbf{Decision\tnote{1}}}                                                                                                                                                                                                   &  & \multicolumn{6}{c}{\textbf{Confidence\tnote{2}}}                                                                                                                                                                                                 &  & \multicolumn{6}{c}{\textbf{Time Consumption}}                                                                               \\ \cmidrule(lr){2-7} \cmidrule(lr){9-14} \cmidrule(l){16-21} 
\multirow{-2}{*}{\textbf{Reviewer}} & 1A                       & 1B                                               & 2A                       & 2B                                               & 3A                       & 3B                                               &  & 1A                       & 1B                                               & 2A                       & 2B                                               & 3A                       & 3B                                               &  & 1A     & 1B                             & 2A     & 2B                             & 3A     & 3B                             \\ \midrule
\textbf{\#1}                        
& {\color[HTML]{3E8853} \textbf{1}} 
& \cellcolor[HTML]{D9D9D9}{\color[HTML]{3E8853} \textbf{1}} 
& {\color[HTML]{3E8853} \textbf{1}} 
& \cellcolor[HTML]{D9D9D9}{\color[HTML]{3E8853} \textbf{1}} 
& {\color[HTML]{3E8853} \textbf{1}} 
& \cellcolor[HTML]{D9D9D9}{\color[HTML]{3E8853} \textbf{1}} 
&  
& {\color[HTML]{3E8853} \textbf{5}} 
& \cellcolor[HTML]{D9D9D9}{\color[HTML]{3E8853} \textbf{5}} 
& {\color[HTML]{3E8853} \textbf{5}} 
& \cellcolor[HTML]{D9D9D9}{\color[HTML]{A1B347} \textbf{4}} 
& {\color[HTML]{3E8853} \textbf{5}} 
& \cellcolor[HTML]{D9D9D9}{\color[HTML]{3E8853} \textbf{5}} 
&  
& $3^{\prime}36^{\prime\prime}$ 
& \cellcolor[HTML]{D9D9D9}$1^{\prime}45^{\prime\prime}$ 
& $4^{\prime}14^{\prime\prime}$ 
& \cellcolor[HTML]{D9D9D9}$1^{\prime}08^{\prime\prime}$ 
& $2^{\prime}54^{\prime\prime}$ 
& \cellcolor[HTML]{D9D9D9}$1^{\prime}03^{\prime\prime}$ 
\\
\textbf{\#2}                        
& {\color[HTML]{3E8853} \textbf{1}} 
& \cellcolor[HTML]{D9D9D9}{\color[HTML]{3E8853} \textbf{1}} 
& {\color[HTML]{3E8853} \textbf{1}} 
& \cellcolor[HTML]{D9D9D9}{\color[HTML]{3E8853} \textbf{1}} 
& {\color[HTML]{C00000} \textbf{0}} 
& \cellcolor[HTML]{D9D9D9}{\color[HTML]{3E8853} \textbf{1}} 
&  
& {\color[HTML]{3E8853} \textbf{5}} 
& \cellcolor[HTML]{D9D9D9}{\color[HTML]{3E8853} \textbf{5}} 
& {\color[HTML]{A1B347} \textbf{4}} 
& \cellcolor[HTML]{D9D9D9}{\color[HTML]{3E8853} \textbf{5}} 
& {\color[HTML]{A1B347} \textbf{4}} 
& \cellcolor[HTML]{D9D9D9}{\color[HTML]{838600} \textbf{3}} 
&  
& $1^{\prime}17^{\prime\prime}$ 
& \cellcolor[HTML]{D9D9D9}$0^{\prime}40^{\prime\prime}$ 
& $4^{\prime}59^{\prime\prime}$ 
& \cellcolor[HTML]{D9D9D9}$2^{\prime}31^{\prime\prime}$ 
& $3^{\prime}59^{\prime\prime}$ 
& \cellcolor[HTML]{D9D9D9}$2^{\prime}43^{\prime\prime}$ 
\\
\textbf{\#3}                        
& {\color[HTML]{3E8853} \textbf{1}} 
& \cellcolor[HTML]{D9D9D9}{\color[HTML]{3E8853} \textbf{1}} 
& {\color[HTML]{3E8853} \textbf{1}} 
& \cellcolor[HTML]{D9D9D9}{\color[HTML]{3E8853} \textbf{1}} 
& {\color[HTML]{3E8853} \textbf{1}} 
& \cellcolor[HTML]{D9D9D9}{\color[HTML]{3E8853} \textbf{1}} 
&  
& {\color[HTML]{3E8853} \textbf{5}} 
& \cellcolor[HTML]{D9D9D9}{\color[HTML]{3E8853} \textbf{5}} 
& {\color[HTML]{A1B347} \textbf{4}} 
& \cellcolor[HTML]{D9D9D9}{\color[HTML]{3E8853} \textbf{5}} 
& {\color[HTML]{A1B347} \textbf{4}} 
& \cellcolor[HTML]{D9D9D9}{\color[HTML]{838600} \textbf{3}} 
&  
& $0^{\prime}37^{\prime\prime}$ 
& \cellcolor[HTML]{D9D9D9}$1^{\prime}19^{\prime\prime}$ 
& $4^{\prime}27^{\prime\prime}$ 
& \cellcolor[HTML]{D9D9D9}$2^{\prime}22^{\prime\prime}$ 
& $7^{\prime}45^{\prime\prime}$ 
& \cellcolor[HTML]{D9D9D9}$7^{\prime}04^{\prime\prime}$ 
\\
\textbf{\#4}                        
& {\color[HTML]{3E8853} \textbf{1}} 
& \cellcolor[HTML]{D9D9D9}{\color[HTML]{3E8853} \textbf{1}} 
& {\color[HTML]{3E8853} \textbf{1}} 
& \cellcolor[HTML]{D9D9D9}{\color[HTML]{3E8853} \textbf{1}} 
& {\color[HTML]{3E8853} \textbf{1}} 
& \cellcolor[HTML]{D9D9D9}{\color[HTML]{3E8853} \textbf{1}} 
&  
& {\color[HTML]{3E8853} \textbf{5}} 
& \cellcolor[HTML]{D9D9D9}{\color[HTML]{3E8853} \textbf{5}} 
& {\color[HTML]{3E8853} \textbf{5}} 
& \cellcolor[HTML]{D9D9D9}{\color[HTML]{3E8853} \textbf{5}} 
& {\color[HTML]{3E8853} \textbf{5}} 
& \cellcolor[HTML]{D9D9D9}{\color[HTML]{3E8853} \textbf{5}} 
&  
& $2^{\prime}30^{\prime\prime}$ 
& \cellcolor[HTML]{D9D9D9}$0^{\prime}14^{\prime\prime}$ 
& $2^{\prime}49^{\prime\prime}$ 
& \cellcolor[HTML]{D9D9D9}$0^{\prime}34^{\prime\prime}$ 
& $2^{\prime}47^{\prime\prime}$ 
& \cellcolor[HTML]{D9D9D9}$1^{\prime}49^{\prime\prime}$ 
\\ \bottomrule
\end{tabular}
}
\newline
\raggedright \footnotesize
\textsuperscript{1}\textbf{Decision}: {\color[HTML]{3E8853} \textbf{1}} and {\color[HTML]{C00000} \textbf{0}} denote the review's decision is correct and incorrect, respectively; 
\textsuperscript{2}\textbf{Confidence}: {\color[HTML]{C00000}\textbf{1}}, {\color[HTML]{CC6600}\textbf{2}}, {\color[HTML]{838600}\textbf{3}}, {\color[HTML]{A1B347}\textbf{4}}, {\color[HTML]{3E8853}\textbf{5}} represent reviewer's decision confidence from the least confident to the most confident;
\textsuperscript{3}\protect\colorbox{lightgrayD}{\textit{Grey background}} denotes the current case is applying \sys;
\end{threeparttable}
\end{table*}

Will \explainer-generated explanations introduce decision-making bias in human reviewers, potentially increasing false positive judgments in cheat detection?
To answer this research question, we conduct a user study including four former reviewers and six cases.
In terms of the case selection, we select cheaters with similar cheating sophistications to compare the decision correctness and time efficiency with and without the use of \sys.
1A and 1B are blatant cheaters (1A and 1B are Case C and Case E in the case study); 2A and 2B are disguised cheaters (2A is newly selected, 2B is Case D in the case study); 3A and 3B are seasoned players that are misclassified by \sys, i.e., the false positives (both cases are newly selected).
The case selections are conducted by two experts listed in \autoref{tab:expert_skill_level}.
The selected cases are reviewed by the reviewers following the procedures in \autoref{tab:us_checklist}. 
The reviewer's skill level is listed in \autoref{tab:reviewer_skill_level}.
The user study results on this research question are demonstrated in \autoref{tab:user_study_result}.
Based on the results, we observe that a reviewer without \sys makes an incorrect decision. 
Reviewer's confidences on the false positives might be influenced by \sys (lower confidence on 3B). 
However, their decisions are not misdirected (correct decisions on 3B). 
\sys primarily accelerates the review process by reducing the time consumption for each case. 
Reviewers with lower confidence tend to spend more time reviewing; with \sys, this additional time caused by lower confidence is effectively eliminated, resulting in faster review times compared to cases with higher-confidence reviewers.

\begin{table}[htbp]
\caption{User study reviewers' CS2 skill level, where \textit{5EPlay} \cite{5eplay_home} and \textit{Perfect World} \cite{wanmei_pvp} are the platform (introduced in~\nameref{sec:ethical}) ranks, \textit{Premier} is the official rank. All invited reviewers are the former reviewers of CS:GO Overwatch Investigators \cite{counter_strike_overwatch}.}
\vspace{ - 0.5em}
\label{tab:reviewer_skill_level}
\resizebox{\columnwidth}{!}{%
\begin{tabular}{@{}clc@{}}
\toprule
\textbf{Reviewer} & \textbf{CS2 Skill Level} & \textbf{Playtime (hrs)} \\ \midrule
\textbf{\#1}      & \textit{5EPlay} - \textbf{A}, \textit{Perfect World} - \textbf{S}                 & 3,563                   \\
\textbf{\#2}      & \textit{5EPlay} - \textbf{A}, \textit{Perfect World} - \textbf{A}, \textit{Premier} - \text{18,491}  & 2,020                   \\
\textbf{\#3}      & \textit{5EPlay} - \textbf{A}, \textit{Perfect World} - \textbf{S}, \textit{Premier} - \text{19,486}  & 2,400                   \\
\textbf{\#4}      & \textit{5EPlay} - \textbf{A+}, \textit{Perfect World} - \textbf{S}, \textit{Premier} - \text{20,000}  & 2,917                   \\ \bottomrule
\end{tabular}
}
\end{table}
\section{Discussion}
\label{sec:discussion}
\noindent\textbf{Data Collection Generalizability.}
While most FPS games incorporate replay systems, certain niche titles or those developed by smaller studios may lack this functionality.
However, since \sys only requires pitch and yaw data, which are fundamental parameters that all FPS games must track to build the game (i.e., the control of the in-game avatar's view directions), developers can easily collect them for \sys as input. 
Notably, \sys is game-agnostic, and any FPS game providing temporal pitch and yaw data can leverage the system for subsequent feature extraction, model training, cheat detection, and interpretable analysis.

\noindent\textbf{Reducing and Expediting Review.}
Given the current anti-cheat development, manual review of the positive suspects is inevitable across all FPS games \cite{hawk2024}. 
Given that the in-game paid items are tied to a player’s account, a ban should not be issued solely based on the anti-cheat system decision. 
It is imperative to provide substantial evidence before initiating such an action. 
Therefore, the final decision requires manual review to avoid false bans.
According to the statistics of our partner platform, their in-use proprietary anti-cheat obtains 47.5\% recall and 10.1\% FPR after the manual re-verification on the label.
In contrast, \sys consistently outperforms both the industrial baseline and prior works in \autoref{sec:compare_sota} across all evaluation metrics.
Regarding manual involvement, since \sys reduces the FPR from 10.1\% to 5.9\% (\autoref{tab:feature_ablation}), this corresponds to a 1.71$\times$ reduction in the number of false positives that require manual review, thereby substantially alleviating the human inspection workload.
Additionally, \sys shortens the review time on cheating samples by providing explainable analyses.
Regarding the ability to identify cheaters, \sys's recall excels among all tested and reported anti-cheat methods.
Considering the worrying performance of the current anti-cheat in most FPS games \cite{zleague_warzone_anticheat, pushtotalk_anticheat, algshack2024, hawk2024, 10.1145/3689934.3690816}, \sys is able to significantly reduce the FNR and FPR at the same time.

\noindent\textbf{Undetectable Cases.}
\sys is designed around elimination events and therefore cannot detect players who do not secure kills. 
For example, purely support-oriented heroes, such as \textit{Mercy} in Overwatch, may not contribute to eliminations, making \sys inapplicable for determining whether they are cheating. 
However, since reducing the opposing team’s player count is a necessary step toward victory in FPS games, cheating without contributing to eliminations has a comparatively limited impact on overall gameplay outcomes. 
In addition, if a cheater employs techniques that do not influence the aiming trajectory (e.g., temporarily toggling wallhacks during encounters, illegally sharing information with teammates), \sys is not applicable as such behaviors do not affect the player’s aiming trajectory; nevertheless, the vast majority of cheaters do not rely on these atypical strategies. 
Regarding genre applicability, \sys is theoretically applicable to any FPS game whose engine uses a pitch–yaw representation for avatar navigation to detect aim-assist–based cheats. 
If a game does not employ a pitch–yaw representation, \sys cannot be directly applied; however, it can still be adapted if an alternative means of obtaining the player’s aiming trajectory is available, by modifying the coordinate transformation in \autoref{alg:py2xy}. 
\section{Related Works}
\label{sec:related_works}
\noindent\textbf{Statistics-based Detection.}
Yu \textit{et al.} \cite{yu2012statistical,yu2012heuristic} detect aim-assist cheats using two statistical features, cursor acceleration and target lock duration, to distinguish between legitimate players and cheaters via the Kolmogorov-Smirnov test. 
However, the approach requires modifications to the client-side game engine, limiting its generalizability across different games. 
Moreover, their features are aggregated over entire matches rather than temporally fine-grained like \sys's, making them less effective at capturing gameplay details. 
Additionally, since their detection operates solely on the client side, it remains vulnerable to memory tampering.
Liu \textit{et al.} \cite{liu2017detecting} exploit behavioral discrepancies in cheaters, such as unusually low skill levels relative to their performance metrics (e.g., kill counts). 
However, their method struggles with highly skilled or motivated players, such as seasoned gamers carefully exploiting the aim-assist cheats, as well as sophisticated cheats (e.g., Cases A and D in our case study).
Galli \textit{et al.} \cite{galli2011unreal} and Alayed \textit{et al.} \cite{alayed2013ml} propose supervised learning approaches that rely on complex features, such as aiming angles, rather than \sys's more generalizable angular change metric. 
These features require modifications to the server-side game engine. For instance, aiming angles depend on target visibility, which varies with client-specific field-of-view settings or is not directly available for some games, such as PUBG. 
Consequently, these methods suffer from limited generalizability.
Beyond these limitations, none of the aforementioned approaches have been validated on large-scale real-world datasets or have considered interpretability, leaving their practical effectiveness uncertain and unreliable.

\noindent\textbf{Behavioral-based Detection.} 
Server-side \textsc{Hawk} \cite{hawk2024} and client-side \textsc{BotScreen} \cite{choi2023botscreen} are the state-of-the-art solutions for aim-assist cheats detection.
\textsc{BotScreen} \cite{choi2023botscreen} leverages one temporal feature, aiming angles, which is similar to the one in the previous statistical methods \cite{galli2011unreal,alayed2013ml} but in a time-series manner. 
Therefore, the generalizability issue inherently exists.
Besides, it is tested with a small dataset constructed by 14 hired players and implemented in a single game.
Thus, its real-world performance and generalizability remain unclear.
Additionally, it leverages unsupervised learning to detect anomalies with only the aforementioned normal players' data as model input, and is tested with only one open-source aimbot.
Considering the diverse cheating sophistication discussed in \autoref{sec:cheat_type} and demonstrated in the previous case studies, \textsc{BotScreen}'s real-world applicability remains questionable.
\textsc{Hawk} \cite{hawk2024} requires massive game-specific features (e.g., flash bang affection duration on the opponents and teammates, incendiary damages, etc.) as mentioned in \autoref{sec:introduction} and \autoref{sec:compare_sota}.
Therefore, their generalizability is confined to CS:GO, with some features failing to reproduce even on its successor, CS2.
Importantly, its performance is lower compared to \sys.
AntiCheatPT \cite{anticheatpt} leverages LLMs for cheat detection. However, its overhead, generalizability, and deployability on the real game server remain unknown.
Additionally, none of the above works are explainable.

\noindent\textbf{Client-side Prevention.} 
BlackMirror \cite{blackmirror2020} leverages SGX to prevent wallhacks by supplying the GPU with only visible entities based on local predictions. 
However, it focuses solely on prevention without detection capabilities, and is limited by hardware requirements, overhead, and its narrow scope targeting a single cheat modality. 
Invisibility Cloak \cite{invisibilitycloak} thwarts computer-vision-based aimbots by obstructing object detection in screen captures, but remains ineffective against most aim-assist cheats that exploit memory access.

\noindent\textbf{Industrial Proprietary Anti-Cheat.} 
Commercial anti-cheat systems such as VAC \cite{SteamSupport}, Vanguard \cite{RiotGamesInc}, EAC \cite{EasyAC}, and BattlEye \cite{BattlEye} provide both detection and prevention across various games via custom integrations \cite{Silva_2022}.
These systems monitor process privileges, scan DNS caches and cookies for browsing histories and cheat purchase traces \cite{lehtonen2020comparative, SteamSupport}, and utilize kernel-level drivers \cite{pushtotalk_anticheat}.
However, they raise significant security and privacy concerns due to hard drive scanning and root privilege requirements, which risk unauthorized access and data leakage \cite{mikkelsen2017information,9566108,aimlow2020}.
Known vulnerabilities, including RCE exploits \cite{BackEngineering2021} and several CVEs \cite{CVE-2019-16098, CVE-2020-36603, CVE-2021-3437, CVE-2023-38817}, further expose users to system instability and privacy breaches \cite{hawk2024}.
These systems also rely heavily on cheat blacklists and malware signatures \cite{9774028, pushtotalk_anticheat, choi2023botscreen, liu2017detecting}, which are ineffective against novel or obfuscated cheats with altered signatures \cite{choi2023botscreen, liu2017detecting, hawk2024}.
Modern paid cheat services frequently auto-update to evade detection, causing industry anti-cheats to exhibit low recall and lag behind emerging threats \cite{hawk2024}.
Moreover, commercial solutions are embedded pre-release and run during gameplay \cite{pushtotalk_anticheat}, introducing client-side overhead and limiting generalizability.
While these systems often combine multiple detection strategies to cross-verify suspicious activity, this extends the ban cycle \cite{pushtotalk_anticheat}.
Finally, being proprietary and closed-source \cite{pushtotalk_anticheat}, they hinder independent research and comparative evaluations without direct collaboration with game developers.

\noindent\textbf{Explainable Detection in Other Genres.} 
Tao \textit{et al.} proposed a four-pronged explainable design for detecting real money trading in MMORPGs like World of Warcraft \cite{tao_xai_2020}.
The interpretability of their work on FPS games is limited to client-side screenshots' anomaly overlays identification (e.g., cheat's bounding box or UI).
This may raise privacy concerns and can be easily bypassed by managing process privileges or disabling the visible overlays.
Hence, it is less practical for real-world deployment.
\section{Conclusion}
\label{sec:conclusion}
We presented \sys, a generalized, explainable, and efficient server-side anti-cheat system for detecting aim-assist cheats in FPS games.
\sys introduces novel detection features and a hybrid framework to effectively identify and explain cheating behaviors.
%
%
We demonstrate that \sys achieves state-of-the-art performance in reliably detecting aim-assist cheaters on different games.

\section*{Acknowledgments}
\label{sec:acknoledgements}

We thank \textit{5EPlay} and \textit{Lilith Games} for sharing and labeling their raw data and their support for verifying ground truth labels. We appreciate the voluntary reviews from \textit{Mr.K}, \textit{z1an}, \textit{mrh929}, and \textit{Demons} (disclosed with their IDs) on the user study cases. We acknowledge the constructive review comments by the anonymous reviewers from USENIX Security and NDSS.
This work is partially supported by the NSFC for Young Scientists of China (No.62202400) and the RGC for Early Career Scheme (No.27210024). Any opinions, findings, or conclusions in this paper are those of the authors and do not necessarily reflect the views of NSFC and RGC.
\section*{Ethical Considerations}
\label{sec:ethical}
\noindent\textbf{Data Source and Acquisition.} 
All data used in this study were provided by a third-party competitive gaming platform \textit{5EPlay} that hosts CS2 servers independently of \textit{Valve Corporation} (the game’s developer). These servers operate under the platform’s own user agreements and privacy policies, while authentication occurs via Steam. \textit{5EPlay} directly provided the de-identified dataset to the research team, in compliance with their platform policies\footnote{\textit{5EPlay Personal Information Protection Policy}, translated by the authors, original in Simplified Chinese: \href{https://arena.5eplay.com/page/privacy}{https://arena.5eplay.com/page/privacy}.}. No private information or personally identifiable data was disclosed to the authors. Players' re-identifiable information was removed prior to data transfer.

\noindent\textbf{Data Usage Agreement.} 
The dataset is governed by the \textit{5EPlay Personal Information Protection Policy}, which permits sharing de-identified information for academic research purposes conducted in the public interest. While the data usage agreement acknowledged by players was not specific to this research, it explicitly allows the platform to share anonymized gameplay data for research. All data were handled with strict confidentiality and in accordance with applicable legal and ethical standards.  

\noindent\textbf{\explainer Potential Impact.} 
\explainer could theoretically influence reviewers’ decisions, potentially increasing the risk of false decisions. To mitigate this risk, \explainer is strictly a supportive interpretability tool intended to help experienced game masters (GMs) quickly identify anomalous gameplay segments. All decisions are made independently by trained GMs, and bans are only enforced when multiple reviewers reach the same conclusion, minimizing the influence of any single reviewer or algorithm. GMs should be trained to know that \sys can make mistakes. Players affected by false positives retain the right to appeal, triggering review by a separate moderation team.  

\noindent\textbf{IRB Review.} 
This research was reviewed and approved by the Human Research Ethics Committee (Reference Number: EA250702) at the author's affiliated institution. The study design, data acquisition, and handling procedures were deemed compliant with human subjects research standards.  

\section*{Open Science}
\phantomsection
\label{sec:openscience}
We open-source \sys's repository, datasets, and artifacts at \href{https://doi.org/10.5281/zenodo.17845614}{https://doi.org/10.5281/zenodo.17845614}, as well as its raw CS2 data listed in \autoref{tab:raw_data}. 

\begin{table}[htbp]
\centering
\caption{\sys's raw data.}
\vspace{-0.5em}
\label{tab:raw_data}
\resizebox{0.8\columnwidth}{!}{%
\begin{tabular}{cc}
\toprule
\textbf{Part} & \textbf{URL} \\
\midrule
1  & \url{https://doi.org/10.5281/zenodo.17838584} \\
2  & \url{https://doi.org/10.5281/zenodo.17838763} \\
3  & \url{https://doi.org/10.5281/zenodo.17838827} \\
4  & \url{https://doi.org/10.5281/zenodo.17838865} \\
5  & \url{https://doi.org/10.5281/zenodo.17839153} \\
6  & \url{https://doi.org/10.5281/zenodo.17844922} \\
7  & \url{https://doi.org/10.5281/zenodo.17844928} \\
8  & \url{https://doi.org/10.5281/zenodo.17844938} \\
9  & \url{https://doi.org/10.5281/zenodo.17844942} \\
10 & \url{https://doi.org/10.5281/zenodo.17844950} \\
11 & \url{https://doi.org/10.5281/zenodo.17844964} \\
12 & \url{https://doi.org/10.5281/zenodo.17844969} \\
13 & \url{https://doi.org/10.5281/zenodo.17844977} \\
\bottomrule
\end{tabular}
}
\end{table}

\bibliographystyle{plain}
\bibliography{citations}
\appendix
\section{Evaluation Metrics}
\label{appx:metrics}
Accuracy measures the proportion of correct predictions among all predictions, calculated as \autoref{eq:accuracy}.
Precision represents the ratio of true positives to all positive predictions with \autoref{eq:precision}.
Recall quantifies the fraction of actual positives correctly identified as \autoref{eq:recall}.
F1 Score provides a harmonic mean of precision (\autoref{eq:precision}) and recall (\autoref{eq:recall}) through \autoref{eq:f1}.
False Positive Rate (FPR) indicates the proportion of negatives incorrectly classified as positives using \autoref{eq:fpr}.

\begin{equation}
\label{eq:accuracy}
\text{Accuracy} = \frac{TP + TN}{TP + TN + FP + FN}.
\end{equation}
\begin{equation}
\label{eq:precision}
\text{Precision} = \frac{TP}{TP + FP},
\end{equation}
\begin{equation}
\label{eq:recall}
\text{Recall} = \frac{TP}{TP + FN}.
\end{equation}
\begin{equation}
\label{eq:f1}
F1 = 2 \times \frac{\text{Precision} \times \text{Recall}}{\text{Precision} + \text{Recall}}.
\end{equation}
\begin{equation}
\label{eq:fpr}
\text{FPR} = \frac{FP}{FP + TN}.
\end{equation}

Weighted averages adjust for class imbalance by weighting each class's contribution proportionally to its sample size. The weighted average is computed as \autoref{eq:weighted_avg}, where $w_i$ is the number of instances of class $i$ and $\text{Metric}_i$ is the metric value for class $i$. Notably, weighted accuracy (\autoref{eq:accuracy}) equals weighted recall (\autoref{eq:recall}) in multi-class settings because both metrics converge when accounting for class weights. This occurs because true negatives (TN) become negligible in multi-class calculations, and the class-weighting causes the denominators to align, making both metrics effectively measure the proportion of correctly identified instances relative to each class's prevalence. The equality specifically holds for weighted averages and not for macro averages or binary classification scenarios.

\begin{equation}
\label{eq:weighted_avg}
\text{Weighted Avg} = \frac{\sum_{i=1}^{n} w_i \times \text{Metric}_i}{\sum_{i=1}^{n} w_i},
\end{equation}
\section{Model Parameters}
\label{appx:model_param}
The hyperparameters and configurations for the models in \inspector are detailed in \autoref{tab:inspector_model_parameters}, \autoref{tab:inspector_aggregation_model_parameters}, and \autoref{tab:inspector_match_model_parameters}.

\section{Training and Validation Loss}
\label{appx:loss}
\begin{figure}[htbp]
\centering
\includegraphics[width=0.4\textwidth]{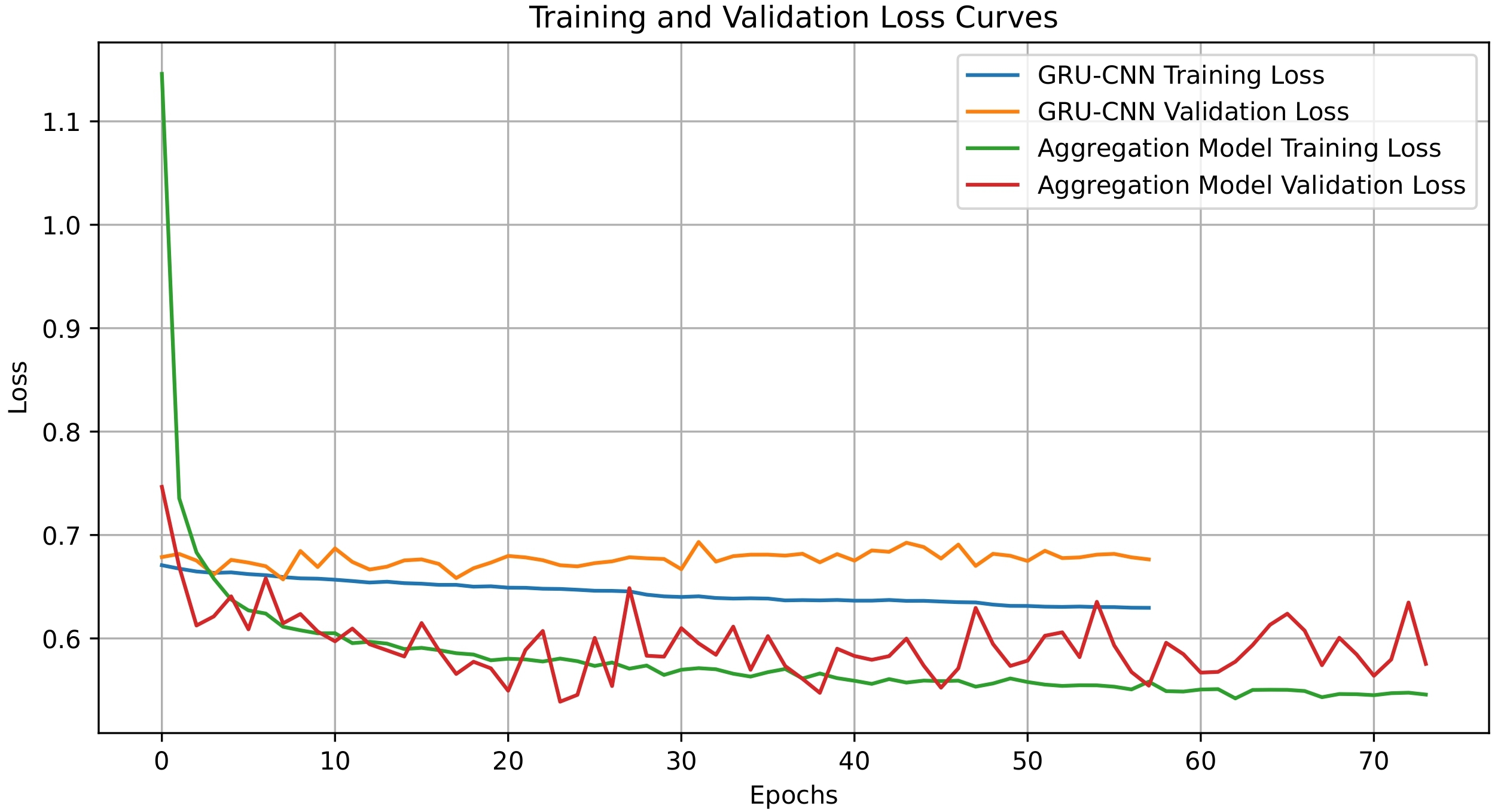}
\caption{Training and validation loss for GRU-CNN and aggregation model.}
\label{fig:loss}
\end{figure}

The training and validation loss are illustrated in \autoref{fig:loss}.

\section{Datasets in Generalizability Evaluation}
\label{appx:generalizability_dataset_description}
The datasets used in generalizability evaluation in \autoref{sec:generalizability} is descripted in \autoref{tab:farlight84_dataset} and \autoref{tab:hawk_dataset}.

\section{Case Study Details}
\label{appx:case_study_addtional}
As discussed in \autoref{sec:cheat_type}, modern aim-assist cheats are highly refined and subtle, making them difficult for novice players to detect. For instance, cheaters in Cases A and D used covert techniques that would likely go unnoticed without expert analysis. Nonetheless, \sys identified their abnormal behaviors and provided interpretable explanations, such as the rapid, unnatural aim adjustments across multiple eliminations between 0:23–0:25 in Case A, and the combination of suspicious actions in Case D, including pre-aiming at unseen enemies, achieving kills with minimal crosshair movement, and quickly flicking away post-shot to mask intent.
In contrast, some cheaters exhibited overt behavior, as in Cases C and E, leaving clear mechanical patterns inconsistent with human play. Examples include unnatural flicking and tracking from 0:42–0:43 in Case C and 0:58 in Case E, as well as implausibly accurate enemy tracking without visual confirmation at 0:46, 0:51, 1:03, and 1:08 in Case C.
Finally, the match-scope case shows how \sys aggregates elimination-level predictions into a match-level decision, visualizing the final verdict by averaging elimination outputs with SHAP values highlighting each feature’s contribution.

\begin{figure}[htbp]
\centering
\includegraphics[width=0.35\textwidth]{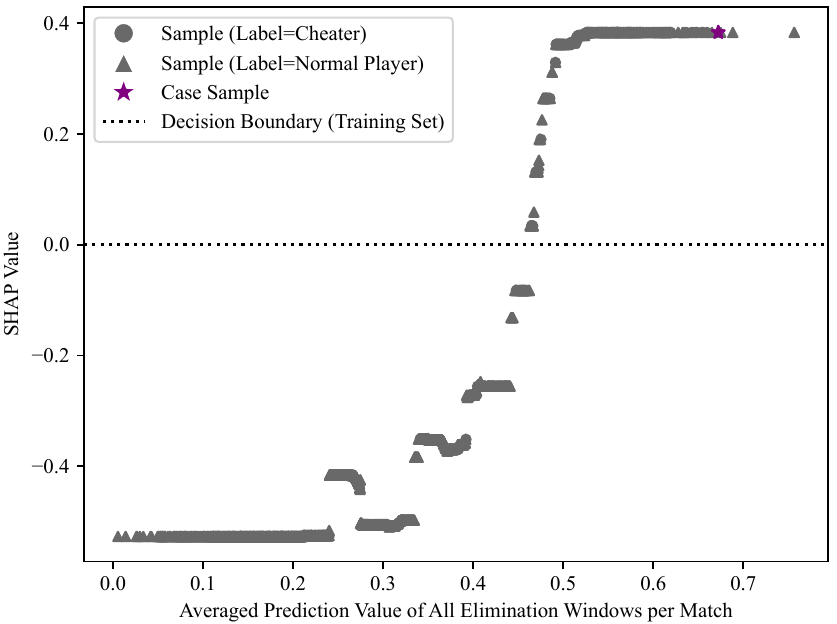}
\caption{Explainable match-scope visualizations of an identified aim-assist cheater.}
\label{fig:case_study_match}
\end{figure}

\begin{figure}[htbp]
\centering
\includegraphics[width=0.4\textwidth]{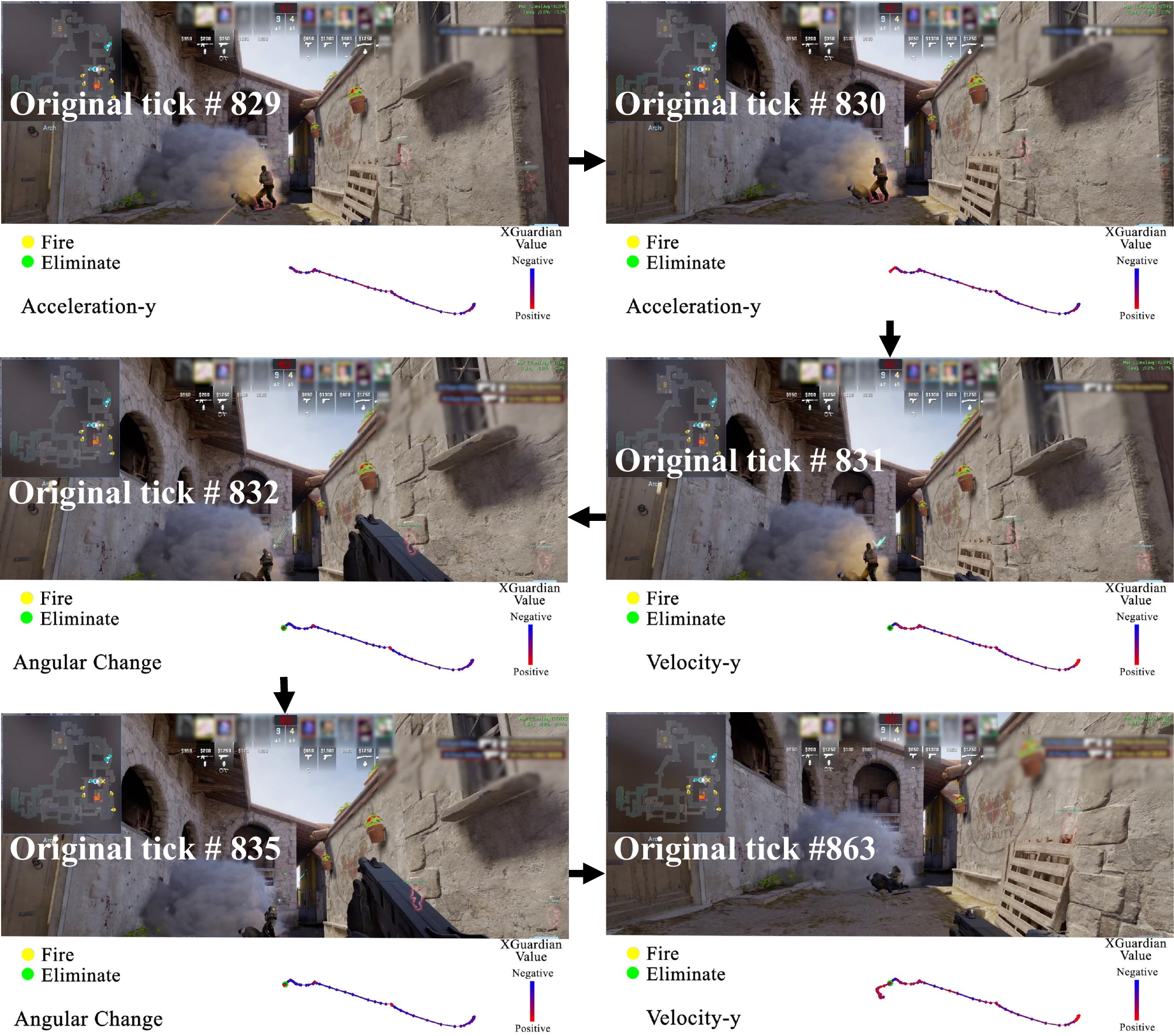}
\caption{Explainable elimination-scope visualizations of an identified aim-assist cheater.}
\label{fig:case_study}
\end{figure}

\autoref{fig:case_study} shows \sys's elimination-level interpretation for Case C’s cheater, selected due to the prominence of their cheating indicators. Within the 7 ticks (109.375 ms) around the elimination, the player’s aiming trajectory exhibits significant anomalies across multiple features, including aimbot-driven oscillations and implausible round-trip travel distances—patterns infeasible for a human in such a brief period.
\autoref{fig:case_study_match} illustrates \sys’s match-level explainable decision-making. The model classifies a sample as a cheater when its input exceeds the learned decision threshold, with the x-axis showing the sample’s input value and the y-axis its SHAP value. A dotted line marks the decision boundary. Other match samples are gray, ground-truth cheaters are circular, legitimate players are triangular, and the highlighted case study is a purple star. Further analysis is available on the our website.

\section{Adversarial Attack Parameter Selection}
\label{appx:adversarial_config}
\autoref{tab:adversarial_configurations} lists different parameters for the adversarial sample selection. 
The sample is selected in a way that a certain feature value is highly similar to the average of the normal player, meanwhile ensuring a relatively large sample size to ensure the results are as unbiased as possible.
The reason for not choosing the deviation in the $\pm15\%$ interval is that the deviation is too large, and it does not meet the requirements for the nature of the sample in adversarial attack.
The reason for not choosing the deviation in the $\pm5\%$ interval is that the sample size was too small and may be prone to bias in the results.

\section{User Study Reviewer's Checklist}
\label{appx:us_review_checklist}
The reviewer's checklist for the user study is listed in \autoref{tab:us_checklist}.

\begin{table}[htbp]
    \centering
    \caption{Embedding generation and trajectory classification stage model architecture. None indicates the variable length of the input batch.}
    \vspace{-0.5em}
    \label{tab:inspector_model_parameters}
    \resizebox{0.75\columnwidth}{!}{%
    \begin{tabular}{lcc}
        \toprule
        \textbf{Type} & \textbf{Output Shape} & \textbf{Param \#} \\
        \midrule
        InputLayer & (None, 6, 8) & 0 \\
        GRU & (None, 6, 64) & 14,208 \\
        LayerNormalization & (None, 6, 64) & 128 \\
        Dropout & (None, 6, 64) & 0 \\
        GRU & (None, 6, 32) & 9,408 \\
        LayerNormalization & (None, 6, 32) & 64 \\
        Dropout & (None, 6, 32) & 0 \\
        Conv1D & (None, 6, 64) & 6,208 \\
        BatchNormalization & (None, 6, 64) & 256 \\
        ReLU & (None, 6, 64) & 0 \\
        Conv1D & (None, 6, 64) & 12,352 \\
        BatchNormalization & (None, 6, 64) & 256 \\
        ReLU & (None, 6, 64) & 0 \\
        Conv1D & (None, 6, 64) & 12,352 \\
        BatchNormalization & (None, 6, 64) & 256 \\
        ReLU & (None, 6, 64) & 0 \\
        GlobalAveragePooling1D & (None, 64) & 0 \\
        Dense & (None, 32) & 2,080 \\
        Dropout & (None, 32) & 0 \\
        Dense & (None, 1) & 33 \\
        \midrule
        \textbf{Total params:} & \multicolumn{2}{c}{57,601 (225.00 KB)} \\
        \textbf{Trainable params:} & \multicolumn{2}{c}{57,217 (223.50 KB)} \\
        \textbf{Non-trainable params:} & \multicolumn{2}{c}{384 (1.50 KB)} \\
        \bottomrule
    \end{tabular}
    }
\end{table}

\begin{table}[htbp]
    \centering
    \footnotesize
    \caption{Aggregation stage model architecture. None indicates the variable length of the input batch.}
    \vspace{-0.5em}
    \label{tab:inspector_aggregation_model_parameters}
    \resizebox{0.75\columnwidth}{!}{%
    \begin{tabular}{lcc}
        \toprule
        \textbf{Type} & \textbf{Output Shape} & \textbf{Param \#} \\
        \midrule
        Dense & (None, 32) & 6,304 \\
        Dropout & (None, 32) & 0 \\
        Dense & (None, 16) & 528 \\
        Dropout & (None, 16) & 0 \\
        Dense & (None, 1) & 17 \\
        \midrule
        \textbf{Total params:} & \multicolumn{2}{c}{6,849 (26.75 KB)} \\
        \textbf{Trainable params:} & \multicolumn{2}{c}{6,849 (26.75 KB)} \\
        \textbf{Non-trainable params:} & \multicolumn{2}{c}{0 (0.00 B)} \\
        \bottomrule
    \end{tabular}
    }
\end{table}

\begin{table}[htbp]
    \centering
    \footnotesize
    \caption{Match prediction stage random forest classifier \cite{scikit-learn_randomforestclassifier} parameters.}
    \label{tab:inspector_match_model_parameters}
    \vspace{-0.5em}
    \resizebox{0.75\columnwidth}{!}{%
    \begin{tabular}{lc}
        \toprule
        \textbf{Parameter} & \textbf{Value} \\
        \midrule
        bootstrap & True \\
        ccp\_alpha & 0.0 \\
        class\_weight & \{0: 0.5316, 1: 8.4213\} \\
        criterion & gini \\
        max\_depth & None \\
        max\_features & sqrt \\
        max\_leaf\_nodes & None \\
        max\_samples & None \\
        min\_impurity\_decrease & 0.0 \\
        min\_samples\_leaf & 100 \\
        min\_samples\_split & 10 \\
        min\_weight\_fraction\_leaf & 0.0 \\
        monotonic\_cst & None \\
        n\_estimators & 100 \\
        n\_jobs & None \\
        oob\_score & False \\
        random\_state & None \\
        verbose & 0 \\
        warm\_start & False \\
        \bottomrule
    \end{tabular}
    }
\end{table}
\begin{table}[htbp]
\centering
\caption{Different parameters for pre-filtering the candidate adversarial samples.}
\label{tab:adversarial_configurations}
\vspace{-0.5em}
\resizebox{0.9\columnwidth}{!}{%
\begin{tabular}{@{}lll@{}}
\toprule
\textbf{Non-cheater deviation} & \textbf{\#Feature meet the deviation} & \textbf{\#Compliant sample} \\ \midrule
\multirow{2}{*}{$\pm5\%$} & $\geq1$ & 9 \\
 & $\geq2$ & None \\
\multirow{2}{*}{\textbf{$\pm10\%$}} & \textbf{$\geq1$} & 22 \\
 & $\geq2$ & None \\
\multirow{3}{*}{$\pm15\%$} & $\geq1$ & 30 \\
 & $\geq2$ & 1 \\
 & $\geq3$ & None \\ \bottomrule
\end{tabular}%
}
\end{table}
\begin{table}[htbp]
\centering
\caption{\textsc{Farlight84} dataset descriptions.}
\label{tab:farlight84_dataset}
\vspace{-0.5em}
\resizebox{\columnwidth}{!}{%
\begin{tabular}{@{}lllllllll@{}}
\toprule
\multirow{2}{*}{\textbf{Scale}} & \multicolumn{2}{c}{\textbf{Train}} &  & \multicolumn{2}{c}{\textbf{Validation}} &  & \multicolumn{2}{c}{\textbf{Test}} \\ \cmidrule(lr){2-3} \cmidrule(lr){5-6} \cmidrule(l){8-9} 
 & \#Normal & \#Cheater &  & \#Normal & \#Cheater &  & \#Normal & \#Cheater \\ \midrule
\textbf{Sample} & 11,402 & 2,601 &  & 8,552 & 1,951 &  & 8,553 & 1,952 \\
\textbf{Match} & 7,554 & 1,493 &  & 6,147 & 1,232 &  & 6,149 & 1,256 \\
\textbf{Player} & 5,513 & 64 &  & 4,678 & 64 &  & 4,691 & 66 \\ \bottomrule
\end{tabular}%
}
\end{table}

\begin{table}[htbp]
\centering
\caption{\textsc{Hawk} dataset descriptions.}
\label{tab:hawk_dataset}
\vspace{-0.5em}
\resizebox{\columnwidth}{!}{%
\begin{tabular}{@{}lllllllll@{}}
\toprule
\multirow{2}{*}{\textbf{Scale}} & \multicolumn{2}{c}{\textbf{Train}} &  & \multicolumn{2}{c}{\textbf{Validation}} &  & \multicolumn{2}{c}{\textbf{Test}} \\ \cmidrule(lr){2-3} \cmidrule(lr){5-6} \cmidrule(l){8-9} 
 & \#Normal & \#Cheater &  & \#Normal & \#Cheater &  & \#Normal & \#Cheater \\ \midrule
\textbf{Sample} & 152,113 & 12,273 &  & 146,543 & 11,970 &  & 146,887 & 12,313 \\
\textbf{Match} & 1,007 & 539 &  & 975 & 528 &  & 993 & 541 \\
\textbf{Player} & 8,971 & 357 &  & 8,660 & 336 &  & 8,800 & 336 \\ \bottomrule
\end{tabular}%
}
\end{table}
\newcommand{\cbox}{$\square$}       
\newcommand{\ccheck}{$\checkmark$}  

\setlist[enumerate]{topsep=0pt, itemsep=1pt, partopsep=0pt, parsep=0pt}

\begin{table}[htbp]
\centering
\caption{User study reviewers' checklist.}
\label{tab:us_checklist}
\vspace{-0.5em}
\small  
\begin{tabular}{p{0.3\columnwidth}|p{0.6\columnwidth}}
\toprule
\textbf{Cases end with A} & \textbf{Cases end with B} \\
\midrule
\begin{minipage}[t]{\linewidth}
\begin{enumerate}[leftmargin=1em,label=\cbox]
    \item Open the demo
    \item Start the timer
    \item Begin the evaluation
    \item Make a decision
    \item Stop the timer
    \item Complete the questionnaire
\end{enumerate}
\end{minipage}
&
\begin{minipage}[t]{\linewidth}
\begin{enumerate}[leftmargin=1em,label=\cbox]
    \item Open the demo
    \item Open PDFs in \texttt{vis} subfolder
    \item Start the timer
    \item Review abnormal segments in PDFs
    \item Begin the evaluation
    \item If the information is sufficient:
    \begin{enumerate}[label=\cbox, leftmargin=1em]
        \item Make a decision
    \end{enumerate}
    \item If the information is insufficient:
    \begin{enumerate}[label=\cbox, leftmargin=1em]
        \item Continue evaluating by yourself until you make a decision
    \end{enumerate}
    \item Compare with the \sys's decision in \texttt{result.txt}
    \begin{enumerate}[label=\cbox, leftmargin=1em]
        \item If different, you may:
        \begin{enumerate}[label=\cbox, leftmargin=1em]
            \item Keep your decision
            \item Re-evaluate, change the decision
        \end{enumerate}
    \end{enumerate}
    \item Stop the timer
    \item Complete the questionnaire
\end{enumerate}
\end{minipage} \\
\bottomrule
\end{tabular}
\end{table}

\end{document}



\newcommand\CQ[1]{\textbf{\textcolor{purple}{CQ: #1}}}
\newcommand\Jiayi[1]{\textbf{\textcolor{orange}{Jiayi: #1}}}

\newcommand{\sys}{\textsc{XGuardian}\xspace}
\newcommand{\inspector}{\textsc{Inspector}\xspace}
\newcommand{\explainer}{\textsc{Explainer}\xspace}

\date{}

\title{USENIX Security '26 Artifact Appendix \\ \sys: Towards Generalized, Explainable and More Effective Server-side Anti-cheat in First-Person Shooter Games}

\author{
{\rm Jiayi Zhang}\\
The University of Hong Kong\\
brucejiayi@connect.hku.hk
\and
{\rm  Chenxin Sun}\\
The University of Hong Kong\\
roniny@connect.hku.hk
\and
{\rm Chenxiong Qian}\\
The University of Hong Kong\\
cqian@cs.hku.hk
} 

\maketitle


\appendix
\section{Artifact Appendix}
This is the artifact evaluation document for \sys, which is going to appear at USENIX Security '26.

\subsection{Abstract}
You may follow the instructions below to access, reuse, and reproduce \sys as well as the experiments mentioned in our paper.

\subsection{Description \& Requirements}

\subsubsection{Security, privacy, and ethical concerns}
\noindent\textbf{Data Source and Acquisition.} 
All data used in this study were provided by a third-party competitive gaming platform \textit{5EPlay} that hosts CS2 servers independently of \textit{Valve Corporation} (the game’s developer). These servers operate under the platform’s own user agreements and privacy policies, while authentication occurs via Steam. \textit{5EPlay} directly provided the de-identified dataset to the research team, in compliance with their platform policies\footnote{\textit{5EPlay Personal Information Protection Policy}, translated by the authors, original in Simplified Chinese: \href{https://arena.5eplay.com/page/privacy}{https://arena.5eplay.com/page/privacy}.}. No private information or personally identifiable data was disclosed to the authors. Players' re-identifiable information was removed prior to data transfer.

\noindent\textbf{Data Usage Agreement.} 
The dataset is governed by the \textit{5EPlay Personal Information Protection Policy}, which permits sharing de-identified information for academic research purposes conducted in the public interest. While the data usage agreement acknowledged by players was not specific to this research, it explicitly allows the platform to share anonymized gameplay data for research. All data were handled with strict confidentiality and in accordance with applicable legal and ethical standards.  

\noindent\textbf{\explainer Potential Impact.} 
\explainer could theoretically influence reviewers’ decisions, potentially increasing the risk of false decisions. To mitigate this risk, \explainer is strictly a supportive interpretability tool intended to help experienced game masters (GMs) quickly identify anomalous gameplay segments. All decisions are made independently by trained GMs, and bans are only enforced when multiple reviewers reach the same conclusion, minimizing the influence of any single reviewer or algorithm. GMs should be trained to know that \sys can make mistakes. Players affected by false positives retain the right to appeal, triggering review by a separate moderation team.  

\noindent\textbf{IRB Review.} 
This research was reviewed and approved by the Human Research Ethics Committee (Reference Number: EA250702) at the author's affiliated institution. The study design, data acquisition, and handling procedures were deemed compliant with human subjects research standards.  

\subsubsection{How to access}
\label{sec:access}
We open-source \sys's repository, datasets, and artifacts at \href{https://doi.org/10.5281/zenodo.17845614}{https://doi.org/10.5281/zenodo.17845614}, as well as its raw CS2 data listed in \autoref{tab:raw_data}. 

\begin{table}[htbp]
\centering
\caption{\sys's raw data.}
\vspace{-0.5em}
\label{tab:raw_data}
\resizebox{0.8\columnwidth}{!}{%
\begin{tabular}{cc}
\toprule
\textbf{Part} & \textbf{URL} \\
\midrule
1  & \url{https://doi.org/10.5281/zenodo.17838584} \\
2  & \url{https://doi.org/10.5281/zenodo.17838763} \\
3  & \url{https://doi.org/10.5281/zenodo.17838827} \\
4  & \url{https://doi.org/10.5281/zenodo.17838865} \\
5  & \url{https://doi.org/10.5281/zenodo.17839153} \\
6  & \url{https://doi.org/10.5281/zenodo.17844922} \\
7  & \url{https://doi.org/10.5281/zenodo.17844928} \\
8  & \url{https://doi.org/10.5281/zenodo.17844938} \\
9  & \url{https://doi.org/10.5281/zenodo.17844942} \\
10 & \url{https://doi.org/10.5281/zenodo.17844950} \\
11 & \url{https://doi.org/10.5281/zenodo.17844964} \\
12 & \url{https://doi.org/10.5281/zenodo.17844969} \\
13 & \url{https://doi.org/10.5281/zenodo.17844977} \\
\bottomrule
\end{tabular}
}
\end{table}

\subsubsection{Hardware dependencies}
Technically, any PC with CPU is able to reuse \sys directly, and any PC with NVIDIA GPU(s) can train and reproduce \sys.
In terms of the experimental configurations, \sys is developed and tested on two different servers. \sys's preprocessing is implemented on an Ubuntu server equipped with 128 cores AMD EPYC 7713P 64-Core Processor and 256 GB RAM. 
The rest of the \sys's experiments including training, validation, testing, metric calculations, and other server-side experiments are conducted on an Ubuntu server equipped with a 24-core Intel Xeon Gold 5418Y CPUs, 503 GB RAM and eight NVIDIA GeForce RTX 4090 24GB (only used two of them for the training).

\subsubsection{Software dependencies}
\textsc{XGuardian} uses go version \texttt{go1.22.5 linux/amd64} for preprocessing. 
The Go compiler can be downloaded from \href{https://go.dev/dl/}{https://go.dev/dl/}. 
Before parsing, use the following command to ensure that \texttt{go} is available in your environment.
\begin{lstlisting}
$ export PATH=$PATH:/usr/local/go/bin
\end{lstlisting}

\noindent\textsc{XGuardian} uses NVIDIA-SMI \texttt{570.124.06}, driver version
\texttt{570.124.06}, CUDA \texttt{12.8}, TensorFlow \texttt{2.17.0},
and cuDNN \texttt{8} for training \sys.
Similar to all ML-based solutions, these are the basic prerequisites for an ML-based project.
This project uses \texttt{conda} to manage packages and environments.
This project is implemented in Python (\href{https://www.python.org/downloads/}{https://www.python.org/downloads/}) version \texttt{3.11.8}.

\subsubsection{Benchmarks}
Three datasets are required: \textsc{CS2} (main dataset used for most of the experiments), \textsc{Farlight84} and \textsc{Hawk} (two sets are used for the comparision and generalizability experiments). All datasets are included in \autoref{sec:access}.

\subsection{Set-up}

Create a Conda environment for preprocessing. Replace the \texttt{prefix} field in \texttt{preprocessing\_clean.yml} with your local \texttt{conda} path.
\begin{lstlisting}
$ conda env create -f XGuardian/XGuardian/env/preprocessing_env/preprocessing_clean.yml
$ conda activate XGuardian-preprocessing
$ pip install -r XGuardian/XGuardian/env/preprocessing_env/preprocessing_pip_requirements.txt
\end{lstlisting}

\noindent Create a detector environment. Replace the \texttt{prefix} field in \texttt{detector\_clean.yml} with your local \texttt{conda} path.
\begin{lstlisting}
$ conda env create -f XGuardian/XGuardian/env/detector/detector_clean.yml
$ conda activate detector_clean
$ pip install -r XGuardian/XGuardian/env/detector/detector_pip_requirements.txt
\end{lstlisting}

\noindent Create an explainer environment. Replace the \texttt{prefix} field in \texttt{explainer\_clean.yml} with your local \texttt{conda} path.
\begin{lstlisting}
$ conda env create -f XGuardian/XGuardian/env/explainer/explainer_clean.yml
$ conda activate explainer_clean
$ pip install -r XGuardian/XGuardian/env/explainer/explainer_pip_requirements.txt
\end{lstlisting}

\noindent \textbf{[\textit{Troubleshooting}]} We use both \texttt{conda \allowbreak install} and \texttt{pip \allowbreak install}. When exporting the environment, \texttt{pip \allowbreak freeze} might capture the absolute path of the packages instead of their version. Might encounter an exception below.

\begin{lstlisting}
ERROR: Could not install packages due to an OSError: [Errno 2] No such file or directory: /croot/xxx
\end{lstlisting}

\noindent To solve this problem, when a line is inside \texttt{requirement.txt} that contains \texttt{@ file:///croot}, discard everything starting from \texttt{@}. For example, below. Alternatively, you may download our revised version of the affected environments from: \href{https://doi.org/10.5281/zenodo.18194986}{https://doi.org/10.5281/zenodo.18194986}.

\begin{lstlisting}
Bottleneck @file:///croot/bottleneck_1707864210935/work --> Bottleneck
\end{lstlisting}


\subsection{Preprocessing}




The raw data can be found at Zendo (please check \autoref{tab:raw_data} above). The raw data is \textbf{EXTREMELY} large, we strongly recommend you to use our processed data (can be found in \texttt{XGuardian/XGuardian/Design/dataset}). \textbf{If you would like to use our processed dataset directly, please skip this entire subsection}.
\subsubsection{Parsing}
Download the raw data and unzip the files.
\begin{lstlisting}
$ cat XGuardian_raw.tar.zst.part.* > XGuardian_raw.tar.zst
$ tar -I zstd -xf XGuardian_raw.tar.zst
\end{lstlisting}

\noindent The initial file structure should be prepared as follows.
\begin{lstlisting}
.
`-- demo_dir/
    |-- demo_0.zip
    |-- demo_1.zip
    |-- ...
    `-- demo_x.zip
\end{lstlisting}

\noindent We use the absolute path when zipping the artifacts. You should rename the path and delete \texttt{home2\allowbreak/data1\allowbreak/J.Zhang/} ahead of the \texttt{XGuardian\allowbreak/XGuardian/}. Activate the conda environment \texttt{XGuradian\allowbreak-preprocessing}. To process concurrently, partition the demos first. You need to enter the demos' root directory \texttt{./demo\_dir} in the command line.
\begin{lstlisting}
$ python partition_creation.py
\end{lstlisting}

\noindent The file structure will then be partitioned as follows.

\begin{lstlisting}
.
`-- demo_dir/
    |-- partition_0/
    |   |-- demo_0.zip
    |   |-- demo_1.zip
    |   |-- ...
    |   `-- demo_x.zip
    |-- ...
    `-- partition_y/
        |-- demo_x+n.zip
        |-- demo_x+n+1.zip
        |-- ...
        `-- demo_x+n+m.zip
\end{lstlisting}

\noindent To parse the demos concurrently, simply run \texttt{shell} script as follows. Note that you should manually modify the number of partitions in the \texttt{for i in \{0..y\}} loop, where \texttt{y} represents the total number of partitions, the \texttt{SCRIPT\_PATH} where you store the source codes, and the \texttt{BASE\_DIR} where it is the demos' root directory.
\begin{lstlisting}
$ chmod +x run_preprocessing_abwh.sh
$ ./run_preprocessing_abwh.sh
\end{lstlisting}

\noindent If processing considerable demos, it will take a long time. After the parsing, the file structure should look like this. The \texttt{log.txt} under each partition folder contains error messages while parsing the demos.
\begin{lstlisting}
.
`-- demo_dir/
    |-- partition_0/
    |   |-- demo_0/
    |   |   |-- demo_0_bomb.csv
    |   |   |-- demo_0_damages.csv
    |   |   |-- demo_0_header.txt
    |   |   |-- demo_0_infernos.csv
    |   |   |-- demo_0_kills.csv
    |   |   |-- demo_0_smokes.csv
    |   |   |-- demo_0_weapon_fires.csv
    |   |   |-- demo_0.dem
    |   |   `-- demo_0.zip
    |   |-- ...
    |   |-- demo_x/
    |   |   |-- demo_x_bomb.csv
    |   |   |-- demo_x_damages.csv
    |   |   |-- demo_x_header.txt
    |   |   |-- demo_x_infernos.csv
    |   |   |-- demo_x_kills.csv
    |   |   |-- demo_x_smokes.csv
    |   |   |-- demo_x_weapon_fires.csv
    |   |   |-- demo_x.dem
    |   |   `-- demo_x.zip
    |   `-- log.txt
    |-- ...
    `-- partition_y/
        |-- demo_x+n/
        |   |-- demo_x+n_bomb.csv
        |   |-- demo_x+n_damages.csv
        |   |-- demo_x+n_header.txt
        |   |-- demo_x+n_infernos.csv
        |   |-- demo_x+n_kills.csv
        |   |-- demo_x+n_smokes.csv
        |   |-- demo_x+n_weapon_fires.csv
        |   |-- demo_x+n.dem
        |   `-- demo_x+n.zip
        |-- ...
        |-- demo_x+n+m/
        |   |-- demo_x+n+m_bomb.csv
        |   |-- demo_x+n+m_damages.csv
        |   |-- demo_x+n+m_header.txt
        |   |-- demo_x+n+m_infernos.csv
        |   |-- demo_x+n+m_kills.csv
        |   |-- demo_x+n+m_smokes.csv
        |   |-- demo_x+n+m_weapon_fires.csv
        |   |-- demo_x+n+m.dem
        |   `-- demo_x+n+m.zip
        `-- log.txt
\end{lstlisting}

\noindent Run the following script to check the integrity and filter the redundant data. The redundant data will be moved to a new directory under the root directory named \texttt{redundant}. If exceptions occurred, detailed logs will be presented in \texttt{integrity\_check\_log.txt}.

\begin{lstlisting}
$ python after_parse_itegrity_check.py <DEMO_ROOT_DIR>
\end{lstlisting}

\noindent The file structure should then look like this.

\begin{lstlisting}
.
`-- demo_dir/
    |-- partition_0/
    |   |-- demo_0/
    |   |   |-- demo_0_bomb.csv
    |   |   |-- demo_0_damages.csv
    |   |   |-- demo_0_header.txt
    |   |   |-- demo_0_infernos.csv
    |   |   |-- demo_0_kills.csv
    |   |   |-- demo_0_smokes.csv
    |   |   |-- demo_0_weapon_fires.csv
    |   |   |-- demo_0.dem
    |   |   `-- demo_0.zip
    |   |-- ...
    |   `-- log.txt
    |-- ...
    |-- partition_y/
    |   |-- demo_x+n/
    |   |   |-- demo_x+n_bomb.csv
    |   |   |-- demo_x+n_damages.csv
    |   |   |-- demo_x+n_header.txt
    |   |   |-- demo_x+n_infernos.csv
    |   |   |-- demo_x+n_kills.csv
    |   |   |-- demo_x+n_smokes.csv
    |   |   |-- demo_x+n_weapon_fires.csv
    |   |   |-- demo_x+n.dem
    |   |   `-- demo_x+n.zip
    |   |-- ...
    |   |-- demo_x+n+m/
    |   |   |-- demo_x+n+m_bomb.csv
    |   |   |-- demo_x+n+m_damages.csv
    |   |   |-- demo_x+n+m_header.txt
    |   |   |-- demo_x+n+m_infernos.csv
    |   |   |-- demo_x+n+m_kills.csv
    |   |   |-- demo_x+n+m_smokes.csv
    |   |   |-- demo_x+n+m_weapon_fires.csv
    |   |   |-- demo_x+n+m.dem
    |   |   `-- demo_x+n+m.zip
    |   `-- log.txt
    `-- redundant/
        `-- .../
            `-- ...
\end{lstlisting}

\noindent Due to enginnering consideration, some redundant folders might be renamed with an addtional suffix \texttt{\_0}, you can run the following scripts to rename them to the original form.

\begin{lstlisting}
$ python redundant_folders_rename.py
> Enter the root directory path: <DEMO_ROOT_DIR>
> [printed info]
> Do you want to proceed with renaming? (y/n): y
> [printed info]
\end{lstlisting}

\noindent Next, label the cheaters within each match. The exceptions will be logged in \texttt{find\_cheater\_log.txt}.

\begin{lstlisting}
$ python Find_Cheaters_in_Folder.py <DEMO_ROOT_DIR> <CHEATER_LIST_EXCEL_DIR>
\end{lstlisting}

\noindent Then your updated file structure should look like this. Each demo directory will now include a \texttt{cheaterlist.csv} file.

\begin{lstlisting}
.
`-- demo_dir/
    |-- partition_0/
    |   |-- demo_0/
    |   |   |-- cheaterlist.csv
    |   |   |-- demo_0_bomb.csv
    |   |   |-- demo_0_damages.csv
    |   |   |-- demo_0_header.txt
    |   |   |-- demo_0_infernos.csv
    |   |   |-- demo_0_kills.csv
    |   |   |-- demo_0_smokes.csv
    |   |   |-- demo_0_weapon_fires.csv
    |   |   |-- demo_0.dem
    |   |   `-- demo_0.zip
    |   |-- ...
    |   `-- log.txt
    |-- ...
    |-- partition_y/
    |   |-- demo_x+n/
    |   |   |-- cheaterlist.csv
    |   |   |-- demo_x+n_bomb.csv
    |   |   |-- demo_x+n_damages.csv
    |   |   |-- demo_x+n_header.txt
    |   |   |-- demo_x+n_infernos.csv
    |   |   |-- demo_x+n_kills.csv
    |   |   |-- demo_x+n_smokes.csv
    |   |   |-- demo_x+n_weapon_fires.csv
    |   |   |-- demo_x+n.dem
    |   |   `-- demo_x+n.zip
    |   |-- ...
    |   |-- demo_x+n+m/
    |   |   |-- cheaterlist.csv
    |   |   |-- demo_x+n+m_bomb.csv
    |   |   |-- demo_x+n+m_damages.csv
    |   |   |-- demo_x+n+m_header.txt
    |   |   |-- demo_x+n+m_infernos.csv
    |   |   |-- demo_x+n+m_kills.csv
    |   |   |-- demo_x+n+m_smokes.csv
    |   |   |-- demo_x+n+m_weapon_fires.csv
    |   |   |-- demo_x+n+m.dem
    |   |   `-- demo_x+n+m.zip
    |   `-- log.txt
    `-- redundant/
        `-- .../
            `-- ...
\end{lstlisting}

\noindent [\textit{\textbf{Optional}}] Some raw zip files might be corrupted during the download, you may use the following script to remove them.

\begin{lstlisting}
$ python remove_corrupt_zips.py <DEMO_ROOT_DIR>
\end{lstlisting}

\subsubsection{Trajectories Extraction}

Run the following \texttt{shell} scripts to concurrently process the trajectory data. You should modify the directory and number of partitions based on your own data.

\begin{lstlisting}
$ chmod +x run_3D_conversion_modified.sh
$ chmod +x run_2D_conversion_modified.sh
$ ./run_3D_conversion_modified.sh
$ ./run_2D_conversion_modified.sh
\end{lstlisting}

\noindent The logs (if any) for 2D/3D trajectory converters will be stored as \texttt{2D\allowbreak\_trajectory\allowbreak\_modified.txt} and \texttt{3D\allowbreak\_trajectory\allowbreak\_modified.txt}, respectively. The file structure should then look like this.

\begin{lstlisting}
.
`-- demo_dir/
    |-- partition_0/
    |   |-- demo_0/
    |   |   |-- 2D_trajectory_modified.csv
    |   |   |-- 3D_trajectory_modified.csv
    |   |   |-- cheaterlist.csv
    |   |   |-- demo_0_bomb.csv
    |   |   |-- demo_0_damages.csv
    |   |   |-- demo_0_header.txt
    |   |   |-- demo_0_infernos.csv
    |   |   |-- demo_0_kills.csv
    |   |   |-- demo_0_smokes.csv
    |   |   |-- demo_0_weapon_fires.csv
    |   |   |-- demo_0.dem
    |   |   `-- demo_0.zip
    |   |-- ...
    |   `-- log.txt
    |-- ...
    |-- partition_y/
    |   |-- demo_x+n/
    |   |   |-- 2D_trajectory_modified.csv
    |   |   |-- 3D_trajectory_modified.csv
    |   |   |-- cheaterlist.csv
    |   |   |-- demo_x+n_bomb.csv
    |   |   |-- demo_x+n_damages.csv
    |   |   |-- demo_x+n_header.txt
    |   |   |-- demo_x+n_infernos.csv
    |   |   |-- demo_x+n_kills.csv
    |   |   |-- demo_x+n_smokes.csv
    |   |   |-- demo_x+n_weapon_fires.csv
    |   |   |-- demo_x+n.dem
    |   |   `-- demo_x+n.zip
    |   |-- ...
    |   |-- demo_x+n+m/
    |   |   |-- 2D_trajectory_modified.csv
    |   |   |-- 3D_trajectory_modified.csv
    |   |   |-- cheaterlist.csv
    |   |   |-- demo_x+n+m_bomb.csv
    |   |   |-- demo_x+n+m_damages.csv
    |   |   |-- demo_x+n+m_header.txt
    |   |   |-- demo_x+n+m_infernos.csv
    |   |   |-- demo_x+n+m_kills.csv
    |   |   |-- demo_x+n+m_smokes.csv
    |   |   |-- demo_x+n+m_weapon_fires.csv
    |   |   |-- demo_x+n+m.dem
    |   |   `-- demo_x+n+m.zip
    |   `-- log.txt
    `-- redundant/
        `-- .../
            `-- ...
\end{lstlisting}

\subsection{\sys General Usage}
\label{sec:detection}

This section describes the general usage of \sys's core detection and explanation components.

\subsubsection{\inspector}

Activate the conda environment \texttt{detector\allowbreak\_clean}. Run \texttt{XGuardian-Inspector.py} with the following command, where \texttt{<train\_path>}, \texttt{<val\_path>}, \texttt{<test\_path>} are the directories of the data splits. According to our paper findings, \texttt{<window\_size>} should be \texttt{6}, \texttt{<stride>} should be \texttt{1}. \texttt{<drop\_columns>} is the parameter to drop some redundant columns in the dataset. This should be set as \texttt{"isCheater,\allowbreak steamid,\allowbreak kill\_num,\allowbreak match\_id,\allowbreak in\_crouch,\allowbreak is\_walking,\allowbreak is\_strafing,\allowbreak weapon,\allowbreak angle,\allowbreak x,\allowbreak y,\allowbreak distance"}. Note that at Line 82-83, the GPUs are designated; please adjust according to your device configurations.
\begin{lstlisting}
$ python XGuardian-Inspector.py <train_path> <val_path> <test_path> <output_dir> <window_size> <stride> <drop_columns>
\end{lstlisting}

\noindent After execution, three CSVs will be generated named as \texttt{aggregated\_preditions\_x.csv}, where x is corresponding to training, validation or testing set, respectively. These CSVs should be first processed with the following commands for some of the following experiments. \texttt{<input\_csv\_path>} should be the directories of the \texttt{aggregated\_preditions\_x.csv}.
\begin{lstlisting}
$ python XGuardian-Inspector-Match-Preprocessing.py <input_csv_path> <output_csv_path>
\end{lstlisting}

\noindent Then run \texttt{XGuardian-Inspector-Match.py} for the final detection result. In which, \texttt{<feature\_selection>} can be set as \texttt{'o'} (original, please use this one normally), \texttt{'b'} (binary), and \texttt{'a'} (all, original+binary). For details, please refer to Section 5.4.5. Note that at Line 46-47, the GPUs are designated; please adjust according to your device configurations.
\begin{lstlisting}
$ python XGuardian-Inspector-Match.py --train_path <training_path> --val_path <validation_path> --test_path <test_path> --feature_set <feature_selection> --output_dir <output_directory>
\end{lstlisting}

\noindent In the output directory, \texttt{round\allowbreak\_integ\allowbreak\_x\allowbreak\_predictions\allowbreak\_n.csv} will be generated, where x is corresponding to training, validation or testing set; n is corresponding to original, binary and all (feature selection). The above CSVs are the final detection results of \sys, you may use \texttt{round\_eval\_metrics\_stats.py} to evaluate the metrics. You should always manually change the directory of \texttt{round\_integ\_x\_predictions\_n.csv} in \texttt{round\_eval\_metrics\_stats.py} before using it.
\begin{lstlisting}
$ python round_eval_metrics_stats.py
\end{lstlisting}

\subsubsection{\explainer}\label{sec:explainer}

Activate the conda environment \texttt{explainer\allowbreak\_clean}. Run \texttt{XGuardian-Explainer-Elimination.py} to generate raw SHAP values per elimination. This script's usage is very similar to \texttt{XGuardian-Inspector.py}. \texttt{<window\_size>} should be \texttt{6}, \texttt{<stride>} should be \texttt{1}. \texttt{<drop\_columns>} is the parameter to drop some redundant columns in the dataset. This should be set as \texttt{"isCheater,\allowbreak steamid,\allowbreak kill\_num,\allowbreak match\_id,\allowbreak in\_crouch,\allowbreak is\_walking,\allowbreak is\_strafing,\allowbreak weapon,\allowbreak angle,\allowbreak x,\allowbreak y,\allowbreak distance"}. This script will not train the detection model again, but will use the trained models (check Line 270), so please make sure the model directories are correct.
\begin{lstlisting}
$ python XGuardian-Explainer.py <train_path> <val_path> <test_path> <output_dir> <window_size> <stride> <drop_columns>
\end{lstlisting}

\noindent The above execution will generate \texttt{shap\allowbreak\_values\allowbreak\_with\allowbreak\_ids.npz}. Then run \texttt{XGuardian-\allowbreak Explainer-\allowbreak Elimination-\allowbreak xguardian\allowbreak\_value\allowbreak\_avg.py} to generate the \sys values per elimination.
You should manually change the input directories: \texttt{csv\allowbreak\_file\allowbreak\_path} (Line 5, your test set directory, i.e., \texttt{Test\allowbreak\_cleaned.csv}) and \texttt{shap\_values} (Line 10, your \texttt{shap\allowbreak\_values\allowbreak\_with\allowbreak\_ids.npz} directory), and \texttt{output\allowbreak\_csv\allowbreak\_file\allowbreak\_path} (Line 81).
\begin{lstlisting}
$ python XGuardian-Explainer-Elimination-xguardian_value_avg.py
\end{lstlisting}

\noindent For elimination visualization, please check \texttt{XGuardian\allowbreak /XGuardian\allowbreak /Design\allowbreak /src\allowbreak /Explainer\allowbreak\_vis} folder.
There are some preprocessing needs to be done. Run \texttt{split\allowbreak\_shap\allowbreak\_csv\allowbreak\_to\allowbreak\_subCSVs.py}. At Line 38-39, you should change the input (the \texttt{XGuardian-\allowbreak Explainer-\allowbreak Elimination-\allowbreak xguardian\allowbreak\_value\allowbreak\_avg.py} generated CSV) and output directories.
\begin{lstlisting}
$ python split_shap_csv_to_subCSVs.py
\end{lstlisting}

\noindent Then it will generate a subfolder \texttt{test\allowbreak\_set\allowbreak\_data}. Run \texttt{2D\allowbreak\_vis.sh} to concurrently visualize the trajectory. In which, at Line 4-5, change the input and output directories. Also, at Line 22, make sure the Python script directory is correct (using \texttt{XGuardian\allowbreak/XGuardian\allowbreak/Design\allowbreak/src\allowbreak/Explainer\allowbreak\_vis\allowbreak/2D\allowbreak\_vis\allowbreak\_batch\allowbreak\_concurrent.py}). At Line 30-31, you should adjust the total number of the cheaters and normal players (check the numbering in \texttt{test\allowbreak\_set\allowbreak\_data}, no need to replace if using our dataset). Due to a different coordination system, the final visualizations (PNGs and MP4s) need to be center-rotated 180 degrees to match the user's aiming trajectories.
\begin{lstlisting}
$ ./2D_vis.sh
\end{lstlisting}

\noindent For match \explainer visualization, run \texttt{XGuardian-\allowbreak Explainer-\allowbreak Match.py}. \texttt{<model\_path>} is the directory of \texttt{logistic\allowbreak\_regression\allowbreak\_model\allowbreak\_original.joblib} (this is the model of Match-Inspector, just ignore the naming, we are actually using random forest).
\begin{lstlisting}
$ python XGuardian-Explainer-Match.py --dataset_path <test_set_directory> --train_path <training_set_directory> --model_path <model_path> --feature_set <feature_selection> --output_dir <output_directory>
\end{lstlisting}

\subsection{Experiments}
This subsection contains the instructions to run and check the raw results of each experiment in our paper.

\subsubsection{Detection Performance}

\textbf{[\textit{To check the raw results}]} Check each \texttt{round\allowbreak\_integ\allowbreak\_test\allowbreak\_predictions\allowbreak\_original.csv} under each subfolder of \texttt{XGuardian\allowbreak/XGuardian\allowbreak/Experiments\allowbreak/5.2\allowbreak\_Detection\allowbreak\_Performace}. Use \texttt{XGuardian\allowbreak/XGuardian\allowbreak/Design\allowbreak/src\allowbreak/round\allowbreak\_eval\allowbreak\_metrics\allowbreak\_stats.py} to calculate the metrics. The directory of the file \texttt{round\allowbreak\_integ\allowbreak\_test\allowbreak\_predictions\allowbreak\_original.csv} must be modified before the execution at Line 5.
\begin{lstlisting}
$ python round_eval_metrics_stats.py
\end{lstlisting}

\noindent\textbf{[\textit{To reproduce}]} Run the following command by replacing the order of the train, validation, and test sets in the dataset thoroughly with the conda environment \texttt{detector\allowbreak\_clean} activated.
\begin{lstlisting}
$ python XGuardian/XGuardian/Design/src/XGuardian-Inspector.py <train_path> <val_path> <test_path> <output_dir> 6 1 "isCheater,steamid,kill_num,match_id,in_crouch,is_walking,is_strafing,weapon,angle,x,y,distance" 
$ python XGuardian/XGuardian/Design/src/XGuardian-Inspector-Match-Preprocessing.py <path_of_each_subfolder>/aggregated_predictions_train.csv <path_of_each_subfolder>/aggregated_predictions_train_cleaned.csv
$ python XGuardian/XGuardian/Design/src/XGuardian-Inspector-Match-Preprocessing.py <path_of_each_subfolder>/aggregated_predictions_val.csv <path_of_each_subfolder>/aggregated_predictions_val_cleaned.csv
$ python XGuardian/XGuardian/Design/src/XGuardian-Inspector-Match-Preprocessing.py <path_of_each_subfolder>/aggregated_predictions_test.csv <path_of_each_subfolder>/aggregated_predictions_test_cleaned.csv
$ python XGuardian-Inspector-Match.py --train_path <path_of_each_subfolder>/aggregated_predictions_train_cleaned.csv --val_path <path_of_each_subfolder>/aggregated_predictions_val_cleaned.csv --test_path <path_of_each_subfolder>/aggregated_predictions_test_cleaned.csv --feature_set 'o' --output_dir <output_directory>
\end{lstlisting}

\subsubsection{Feature Ablation Study}

\textbf{[\textit{To check the raw results}]} Check each \texttt{round\allowbreak\_integ\allowbreak\_test\allowbreak\_predictions\allowbreak\_original.csv} under each subfolder of \texttt{XGuardian\allowbreak/XGuardian\allowbreak/Experiments\allowbreak/5.3\allowbreak\_Feature\allowbreak\_Ablation\allowbreak\_Study\allowbreak/feature\allowbreak\_selection}. Use \texttt{XGuardian\allowbreak/XGuardian\allowbreak/Design\allowbreak/src\allowbreak/round\allowbreak\_eval\allowbreak\_metrics\allowbreak\_stats.py} to calculate the metrics. The directory of the file \texttt{round\allowbreak\_integ\allowbreak\_test\allowbreak\_predictions\allowbreak\_original.csv} must be modified before the execution at Line 5.
\begin{lstlisting}
$ python round_eval_metrics_stats.py
\end{lstlisting}

\noindent\textbf{[\textit{To reproduce}]} Run the following command by modifying the removed columns parameter from the dataset with the conda environment \texttt{detector\allowbreak\_clean} activated. As default, \texttt{<drop\_columns>} would be \texttt{"isCheater,\allowbreak steamid,\allowbreak kill\_num,\allowbreak match\_id,\allowbreak in\_crouch,\allowbreak is\_walking,\allowbreak is\_strafing,\allowbreak weapon,\allowbreak angle,\allowbreak x,\allowbreak y,\allowbreak distance"}. This is the case \texttt{all}, where all of the features are included. When doing the ablation study, one might just add the feature that would like to be ablated in \texttt{<drop\_columns>}. For example, when removing angular change, add \texttt{angle\_change} to the default parameter. Other valid options to be added are \texttt{velocity\_x,\allowbreak velocity\_y,\allowbreak acceleration\_x,\allowbreak acceleration\_y}.
\begin{lstlisting}
$ python XGuardian-Inspector.py XGuardian/XGuardian/Design/dataset/Train_cleaned.csv XGuardian/XGuardian/Design/dataset/Validation_cleaned.csv XGuardian/XGuardian/Design/dataset/Test_cleaned.csv <output_dir> 6 1 <drop_columns> 
$ python XGuardian/XGuardian/Design/src/XGuardian-Inspector-Match-Preprocessing.py <path_of_each_subfolder>/aggregated_predictions_train.csv <path_of_each_subfolder>/aggregated_predictions_train_cleaned.csv
$ python XGuardian/XGuardian/Design/src/XGuardian-Inspector-Match-Preprocessing.py <path_of_each_subfolder>/aggregated_predictions_val.csv <path_of_each_subfolder>/aggregated_predictions_val_cleaned.csv
$ python XGuardian/XGuardian/Design/src/XGuardian-Inspector-Match-Preprocessing.py <path_of_each_subfolder>/aggregated_predictions_test.csv <path_of_each_subfolder>/aggregated_predictions_test_cleaned.csv
$ python XGuardian-Inspector-Match.py --train_path <path_of_each_subfolder>/aggregated_predictions_train_cleaned.csv --val_path <path_of_each_subfolder>/aggregated_predictions_val_cleaned.csv --test_path <path_of_each_subfolder>/aggregated_predictions_test_cleaned.csv --feature_set 'o' --output_dir <output_directory>
\end{lstlisting}

\subsubsection{Elimination Window Comparison}

\textbf{[\textit{To check the raw results}]} Check each \texttt{round\allowbreak\_integ\allowbreak\_test\allowbreak\_predictions\allowbreak\_original.csv} under each subfolder of \texttt{XGuardian\allowbreak/XGuardian\allowbreak/Experiments\allowbreak/5.4.1\allowbreak\_Different\allowbreak\_Elimination\allowbreak\_Window\allowbreak\_Sizes\allowbreak/diff\allowbreak\_window\allowbreak\_comp\allowbreak/results}. Use \texttt{XGuardian\allowbreak/XGuardian\allowbreak/Design\allowbreak/src\allowbreak/round\allowbreak\_eval\allowbreak\_metrics\allowbreak\_stats.py} to calculate the metrics. The directory of the file \texttt{round\allowbreak\_integ\allowbreak\_test\allowbreak\_predictions\allowbreak\_original.csv} must be modified before the execution at Line 5.
\begin{lstlisting}
$ python round_eval_metrics_stats.py
\end{lstlisting}

\noindent\textbf{[\textit{To reproduce}]} Run the following commands with the conda environment \texttt{detector\allowbreak\_clean} activated.
\begin{lstlisting}
$ python XGuardian/XGuardian/Experiments/5.4.1_Different_Elimination_Window_Sizes/diff_window_comp/src/XGuardian_128.py XGuardian/XGuardian/Experiments/5.4.1_Different_Elimination_Window_Sizes/diff_window_comp/ready_for_eval_data/128_train.csv XGuardian/XGuardian/Experiments/5.4.1_Different_Elimination_Window_Sizes/diff_window_comp/ready_for_eval_data/128_val.csv XGuardian/XGuardian/Experiments/5.4.1_Different_Elimination_Window_Sizes/diff_window_comp/ready_for_eval_data/128_test.csv <output_dir> 6 1 "isCheater,steamid,kill_num,match_id,in_crouch,is_walking,is_strafing,weapon,angle,x,y,distance" 
$ python XGuardian/XGuardian/Experiments/5.4.1_Different_Elimination_Window_Sizes/diff_window_comp/src/XGuardian_256.py XGuardian/XGuardian/Experiments/5.4.1_Different_Elimination_Window_Sizes/diff_window_comp/ready_for_eval_data/256_train.csv XGuardian/XGuardian/Experiments/5.4.1_Different_Elimination_Window_Sizes/diff_window_comp/ready_for_eval_data/256_val.csv XGuardian/XGuardian/Experiments/5.4.1_Different_Elimination_Window_Sizes/diff_window_comp/ready_for_eval_data/256_test.csv <output_dir> 6 1 "isCheater,steamid,kill_num,match_id,in_crouch,is_walking,is_strafing,weapon,angle,x,y,distance" 
$ python XGuardian/XGuardian/Design/src/XGuardian-Inspector-Match-Preprocessing.py <path_of_each_subfolder>/aggregated_predictions_train.csv <path_of_each_subfolder>/aggregated_predictions_train_cleaned.csv
$ python XGuardian/XGuardian/Design/src/XGuardian-Inspector-Match-Preprocessing.py <path_of_each_subfolder>/aggregated_predictions_val.csv <path_of_each_subfolder>/aggregated_predictions_val_cleaned.csv
$ python XGuardian/XGuardian/Design/src/XGuardian-Inspector-Match-Preprocessing.py <path_of_each_subfolder>/aggregated_predictions_test.csv <path_of_each_subfolder>/aggregated_predictions_test_cleaned.csv
$ python XGuardian-Inspector-Match.py --train_path <path_of_each_subfolder>/aggregated_predictions_train_cleaned.csv --val_path <path_of_each_subfolder>/aggregated_predictions_val_cleaned.csv --test_path <path_of_each_subfolder>/aggregated_predictions_test_cleaned.csv --feature_set 'o' --output_dir <output_directory>
\end{lstlisting}

\subsubsection{Sliding Window Comparison}
\textbf{[\textit{To check the raw results}]} Check each \texttt{round\allowbreak\_integ\allowbreak\_test\allowbreak\_predictions\allowbreak\_original.csv} under each subfolder of \texttt{XGuardian\allowbreak/XGuardian\allowbreak/Experiments\allowbreak/5.4.2\allowbreak\_Different\allowbreak\_Sliding\allowbreak\_Window\allowbreak\_Sizes\allowbreak/diff\allowbreak\_sliding\allowbreak\_window\allowbreak\_comp}. Use \texttt{XGuardian\allowbreak/XGuardian\allowbreak/Design\allowbreak/src\allowbreak/round\allowbreak\_eval\allowbreak\_metrics\allowbreak\_stats.py} to calculate the metrics. The directory of the file \texttt{round\allowbreak\_integ\allowbreak\_test\allowbreak\_predictions\allowbreak\_original.csv} must be modified before the execution at Line 5.
\begin{lstlisting}
$ python round_eval_metrics_stats.py
\end{lstlisting}

\noindent\textbf{[\textit{To reproduce}]} Run the following commands with the conda environment \texttt{detector\allowbreak\_clean} activated. Modify sliding window parameter \texttt{<window\allowbreak\_size>} accordingly.
\begin{lstlisting}
$ python XGuardian-Inspector.py XGuardian/XGuardian/Design/dataset/Train_cleaned.csv XGuardian/XGuardian/Design/dataset/Validation_cleaned.csv XGuardian/XGuardian/Design/dataset/Test_cleaned.csv <output_dir> <window_size> 1 "isCheater,steamid,kill_num,match_id,in_crouch,is_walking,is_strafing,weapon,angle,x,y,distance" 
$ python XGuardian/XGuardian/Design/src/XGuardian-Inspector-Match-Preprocessing.py <path_of_each_subfolder>/aggregated_predictions_train.csv <path_of_each_subfolder>/aggregated_predictions_train_cleaned.csv
$ python XGuardian/XGuardian/Design/src/XGuardian-Inspector-Match-Preprocessing.py <path_of_each_subfolder>/aggregated_predictions_val.csv <path_of_each_subfolder>/aggregated_predictions_val_cleaned.csv
$ python XGuardian/XGuardian/Design/src/XGuardian-Inspector-Match-Preprocessing.py <path_of_each_subfolder>/aggregated_predictions_test.csv <path_of_each_subfolder>/aggregated_predictions_test_cleaned.csv
$ python XGuardian-Inspector-Match.py --train_path <path_of_each_subfolder>/aggregated_predictions_train_cleaned.csv --val_path <path_of_each_subfolder>/aggregated_predictions_val_cleaned.csv --test_path <path_of_each_subfolder>/aggregated_predictions_test_cleaned.csv --feature_set 'o' --output_dir <output_directory>
\end{lstlisting}

\subsubsection{Different Model Comparison in Embedding Generation}
\textbf{[\textit{To check the raw results}]} Check each \texttt{round\allowbreak\_integ\allowbreak\_test\allowbreak\_predictions\allowbreak\_original.csv} under each subfolder of \texttt{XGuardian\allowbreak/XGuardian\allowbreak/Experiments\allowbreak/5.4.3\allowbreak\_Different\allowbreak\_Models\allowbreak\_in\allowbreak\_Embedding\allowbreak\_Generation\allowbreak/embedding}. Use \texttt{XGuardian\allowbreak/XGuardian\allowbreak/Design\allowbreak/src\allowbreak/round\allowbreak\_eval\allowbreak\_metrics\allowbreak\_stats.py} to calculate the metrics. The directory of the file \texttt{round\allowbreak\_integ\allowbreak\_test\allowbreak\_predictions\allowbreak\_original.csv} must be modified before the execution at Line 5.
\begin{lstlisting}
$ python round_eval_metrics_stats.py
\end{lstlisting}

\noindent\textbf{[\textit{To reproduce}]} BiLSTM and Transformer can be directly executed under \texttt{detector\allowbreak\_clean} environment with the following commands, respectively.
\begin{lstlisting}
$ python XGuardian/XGuardian/Experiments/5.4.3_Different_Models_in_Embedding_Generation/embedding/BiLSTM.py XGuardian/XGuardian/Design/dataset/Train_cleaned.csv XGuardian/XGuardian/Design/dataset/Validation_cleaned.csv XGuardian/XGuardian/Design/dataset/Test_cleaned.csv <output_dir> 6 1 "isCheater,steamid,kill_num,match_id,in_crouch,is_walking,is_strafing,weapon,angle,x,y,distance"
$ python XGuardian/XGuardian/Experiments/5.4.3_Different_Models_in_Embedding_Generation/embedding/Transformer.py XGuardian/XGuardian/Design/dataset/Train_cleaned.csv XGuardian/XGuardian/Design/dataset/Validation_cleaned.csv XGuardian/XGuardian/Design/dataset/Test_cleaned.csv <output_dir> 6 1 "isCheater,steamid,kill_num,match_id,in_crouch,is_walking,is_strafing,weapon,angle,x,y,distance"
$ python XGuardian/XGuardian/Design/src/XGuardian-Inspector-Match-Preprocessing.py <path_of_each_subfolder>/aggregated_predictions_train.csv <path_of_each_subfolder>/aggregated_predictions_train_cleaned.csv
$ python XGuardian/XGuardian/Design/src/XGuardian-Inspector-Match-Preprocessing.py <path_of_each_subfolder>/aggregated_predictions_val.csv <path_of_each_subfolder>/aggregated_predictions_val_cleaned.csv
$ python XGuardian/XGuardian/Design/src/XGuardian-Inspector-Match-Preprocessing.py <path_of_each_subfolder>/aggregated_predictions_test.csv <path_of_each_subfolder>/aggregated_predictions_test_cleaned.csv
$ python XGuardian-Inspector-Match.py --train_path <path_of_each_subfolder>/aggregated_predictions_train_cleaned.csv --val_path <path_of_each_subfolder>/aggregated_predictions_val_cleaned.csv --test_path <path_of_each_subfolder>/aggregated_predictions_test_cleaned.csv --feature_set 'o' --output_dir <output_directory>
\end{lstlisting}

\noindent Clone the environment \texttt{detector\allowbreak\_clean} and install \texttt{keras\allowbreak-tcn} (\href{https://github.com/philipperemy/keras-tcn}{https://github.com/philipperemy/keras-tcn}). A segement of the code (from Line 183 to Line 221 was made as comments during the debugging and testing phases, please cancel the comment hashtags while training).
\begin{lstlisting}
$ conda create --name tcn --clone detector_clean
$ conda activate tcn
$ pip install keras-tcn
$ python XGuardian/XGuardian/Experiments/5.4.3_Different_Models_in_Embedding_Generation/embedding/TCN.py XGuardian/XGuardian/Design/dataset/Train_cleaned.csv XGuardian/XGuardian/Design/dataset/Validation_cleaned.csv XGuardian/XGuardian/Design/dataset/Test_cleaned.csv <output_dir> 6 1 "isCheater,steamid,kill_num,match_id,in_crouch,is_walking,is_strafing,weapon,angle,x,y,distance"
$ python XGuardian/XGuardian/Design/src/XGuardian-Inspector-Match-Preprocessing.py <path_of_each_subfolder>/aggregated_predictions_train.csv <path_of_each_subfolder>/aggregated_predictions_train_cleaned.csv
$ python XGuardian/XGuardian/Design/src/XGuardian-Inspector-Match-Preprocessing.py <path_of_each_subfolder>/aggregated_predictions_val.csv <path_of_each_subfolder>/aggregated_predictions_val_cleaned.csv
$ python XGuardian/XGuardian/Design/src/XGuardian-Inspector-Match-Preprocessing.py <path_of_each_subfolder>/aggregated_predictions_test.csv <path_of_each_subfolder>/aggregated_predictions_test_cleaned.csv
$ python XGuardian-Inspector-Match.py --train_path <path_of_each_subfolder>/aggregated_predictions_train_cleaned.csv --val_path <path_of_each_subfolder>/aggregated_predictions_val_cleaned.csv --test_path <path_of_each_subfolder>/aggregated_predictions_test_cleaned.csv --feature_set 'o' --output_dir <output_directory>
\end{lstlisting}

\subsubsection{Different Model Comparison in Trajectory Classification}
\textbf{[\textit{To check the raw results}]} Check each \texttt{round\allowbreak\_integ\allowbreak\_test\allowbreak\_predictions\allowbreak\_original.csv} under each subfolder of \texttt{XGuardian\allowbreak/XGuardian\allowbreak/Experiments/5.4.4\allowbreak\_Different\allowbreak\_Models\allowbreak\_in\allowbreak\_Trajectory\allowbreak\_Classification\allowbreak/integration}. Use \texttt{XGuardian\allowbreak/XGuardian\allowbreak/Design\allowbreak/src\allowbreak/round\allowbreak\_eval\allowbreak\_metrics\allowbreak\_stats.py} to calculate the metrics. The directory of the file \texttt{round\allowbreak\_integ\allowbreak\_test\allowbreak\_predictions\allowbreak\_original.csv} must be modified before the execution at Line 5.
\begin{lstlisting}
$ python round_eval_metrics_stats.py
\end{lstlisting}

\noindent\textbf{[\textit{To reproduce}]} Run the following command with the conda environment \texttt{detector\allowbreak\_clean} activated.
\begin{lstlisting}
$ python /home2/data1/J.Zhang/XGuardian/XGuardian/Experiments/5.4.4_Different_Models_in_Trajectory_Classification/integration/GRU-fc.py XGuardian/XGuardian/Design/dataset/Train_cleaned.csv XGuardian/XGuardian/Design/dataset/Validation_cleaned.csv XGuardian/XGuardian/Design/dataset/Test_cleaned.csv <output_dir> 6 1 "isCheater,steamid,kill_num,match_id,in_crouch,is_walking,is_strafing,weapon,angle,x,y,distance" 
$ python XGuardian/XGuardian/Design/src/XGuardian-Inspector-Match-Preprocessing.py <path_of_each_subfolder>/aggregated_predictions_train.csv <path_of_each_subfolder>/aggregated_predictions_train_cleaned.csv
$ python XGuardian/XGuardian/Design/src/XGuardian-Inspector-Match-Preprocessing.py <path_of_each_subfolder>/aggregated_predictions_val.csv <path_of_each_subfolder>/aggregated_predictions_val_cleaned.csv
$ python XGuardian/XGuardian/Design/src/XGuardian-Inspector-Match-Preprocessing.py <path_of_each_subfolder>/aggregated_predictions_test.csv <path_of_each_subfolder>/aggregated_predictions_test_cleaned.csv
$ python XGuardian-Inspector-Match.py --train_path <path_of_each_subfolder>/aggregated_predictions_train_cleaned.csv --val_path <path_of_each_subfolder>/aggregated_predictions_val_cleaned.csv --test_path <path_of_each_subfolder>/aggregated_predictions_test_cleaned.csv --feature_set 'o' --output_dir <output_directory>
\end{lstlisting}

\subsubsection{Different Input Style Comparison in Elimination Prediction}
\textbf{[\textit{To check the raw results}]} Check each \texttt{round\allowbreak\_integ\allowbreak\_test\allowbreak\_predictions\allowbreak\_original.csv} under each subfolder of \texttt{XGuardian\allowbreak/XGuardian\allowbreak/Experiments\allowbreak/5.4.5\allowbreak\_Input\allowbreak\_Styles\allowbreak\_in\allowbreak\_Elimination\allowbreak\_Predictions}. Use \texttt{XGuardian\allowbreak/XGuardian\allowbreak/Design\allowbreak/src\allowbreak/round\allowbreak\_eval\allowbreak\_metrics\allowbreak\_stats.py} to calculate the metrics. The directory of the file \texttt{round\allowbreak\_integ\allowbreak\_test\allowbreak\_predictions\allowbreak\_original.csv} must be modified before the execution at Line 5.
\begin{lstlisting}
$ python round_eval_metrics_stats.py
\end{lstlisting}

\noindent\textbf{[\textit{To reproduce}]} Run \texttt{XGuardian-Inspector-Match.py} for the final detection result. In which, \texttt{<feature\_selection>} can be set as \texttt{'o'} (original, please use this one normally), \texttt{'b'} (binary), and \texttt{'a'} (all, original+binary).
\begin{lstlisting}
$ python XGuardian-Inspector-Match.py --train_path XGuardian/XGuardian/Experiments/5.2_Detection_Performace/overall_performance/TrainValTest/aggregated_predictions_train_cleaned.csv --val_path XGuardian/XGuardian/Experiments/5.2_Detection_Performace/overall_performance/TrainValTest/aggregated_predictions_val_cleaned.csv --test_path XGuardian/XGuardian/Experiments/5.2_Detection_Performace/overall_performance/TrainValTest/aggregated_predictions_test_cleaned.csv --feature_set <feature_selection> --output_dir <output_directory>
\end{lstlisting}

\subsubsection{Comparative Study}
\textbf{[\textit{To check the raw results}]} In terms of \textsc{Hawk}. Check each \texttt{round\allowbreak\_integ\allowbreak\_test\allowbreak\_predictions\allowbreak\_original.csv} under each subfolder of \texttt{XGuardian\allowbreak/XGuardian\allowbreak/Experiments/5.5\allowbreak\_Comparative\allowbreak\_Study}. Use \texttt{XGuardian\allowbreak/XGuardian\allowbreak/Design\allowbreak/src\allowbreak/round\allowbreak\_eval\allowbreak\_metrics\allowbreak\_stats.py} to calculate the metrics. The directory of the file \texttt{round\allowbreak\_integ\allowbreak\_test\allowbreak\_predictions\allowbreak\_original.csv} must be modified before the execution at Line 5.
\begin{lstlisting}
$ python round_eval_metrics_stats.py
\end{lstlisting}

\noindent In terms of statistical methods. Run the following commands.
\begin{lstlisting}
$ python XGuardian/XGuardian/Experiments/5.5_Comparative_Study/stats_method/AccA_eval.py
$ python XGuardian/XGuardian/Experiments/5.5_Comparative_Study/stats_method/HitAcc_eval.py
\end{lstlisting}

\noindent\textbf{[\textit{To reproduce}]} In terms of \textsc{Hawk}. Run the following commands. Due to minimizing the size of our artifacts, we deleted some intermediate files (e.g., \texttt{aggregated\allowbreak\_prediction\allowbreak\_x.csv}, etc.), you may reproduce it by using \texttt{XGuardian\allowbreak-Inspector.py}.
\begin{lstlisting}
$ python XGuardian/XGuardian/Experiments/5.5_Comparative_Study/hawk/hawk_xguardian.py XGuardian/XGuardian/Experiments/5.5_Comparative_Study/hawk/HAWK_train.csv XGuardian/XGuardian/Experiments/5.5_Comparative_Study/hawk/HAWK_validation.csv XGuardian/XGuardian/Experiments/5.5_Comparative_Study/hawk/HAWK_test.csv <output_dir> 6 1 "isCheater,steamid,kill_num,match_id,in_crouch,is_walking,is_strafing,weapon,angle,x,y,distance" 
$ python XGuardian/XGuardian/Design/src/XGuardian-Inspector-Match-Preprocessing.py XGuardian/XGuardian/Experiments/5.5_Comparative_Study/hawk/aggregated_predictions_train.csv XGuardian/XGuardian/Experiments/5.5_Comparative_Study/hawk/aggregated_predictions_train_cleaned.csv
$ python XGuardian/XGuardian/Design/src/XGuardian-Inspector-Match-Preprocessing.py XGuardian/XGuardian/Experiments/5.5_Comparative_Study/hawk/aggregated_predictions_val.csv XGuardian/XGuardian/Experiments/5.5_Comparative_Study/hawk/aggregated_predictions_val_cleaned.csv
$ python XGuardian/XGuardian/Design/src/XGuardian-Inspector-Match-Preprocessing.py XGuardian/XGuardian/Experiments/5.5_Comparative_Study/hawk/aggregated_predictions_test.csv XGuardian/XGuardian/Experiments/5.5_Comparative_Study/hawk/aggregated_predictions_test_cleaned.csv
$ python XGuardian-Inspector-Match.py --train_path XGuardian/XGuardian/Experiments/5.5_Comparative_Study/hawk/aggregated_predictions_train_cleaned.csv --val_path XGuardian/XGuardian/Experiments/5.5_Comparative_Study/hawk/aggregated_predictions_val_cleaned.csv --test_path XGuardian/XGuardian/Experiments/5.5_Comparative_Study/hawk/aggregated_predictions_test_cleaned.csv --feature_set 'o' --output_dir <output_directory>
\end{lstlisting}

\noindent In terms of statistical methods. Run the following commands.
\begin{lstlisting}
$ python XGuardian/XGuardian/Experiments/5.5_Comparative_Study/stats_method/AccA_eval.py
$ python XGuardian/XGuardian/Experiments/5.5_Comparative_Study/stats_method/HitAcc_eval.py
\end{lstlisting}

\subsubsection{Generalizability}
\textbf{[\textit{To check the raw results}]} Check each \texttt{round\allowbreak\_integ\allowbreak\_test\allowbreak\_predictions\allowbreak\_original.csv} under each subfolder of \texttt{XGuardian\allowbreak/XGuardian\allowbreak/Experiments\allowbreak/5.6\allowbreak\_Generalizability\allowbreak/farlight84\allowbreak/result\allowbreak/mobile}. Use \texttt{XGuardian\allowbreak/XGuardian\allowbreak/Design\allowbreak/src\allowbreak/round\allowbreak\_eval\allowbreak\_metrics\allowbreak\_stats.py} to calculate the metrics. The directory of the file \texttt{round\allowbreak\_integ\allowbreak\_test\allowbreak\_predictions\allowbreak\_original.csv} must be modified before the execution at Line 5.
\begin{lstlisting}
$ python round_eval_metrics_stats.py
\end{lstlisting}

\noindent\textbf{[\textit{To reproduce}]} Run the following commands. Due to minimizing the size of our artifacts, we deleted some intermediate files (e.g., \texttt{aggregated\allowbreak\_prediction\allowbreak\_x.csv}, etc.), you may reproduce it by using \texttt{XGuardian\allowbreak-Inspector.py}. For \textsc{Hawk}'s results and reproducible procedures, please check the previous subsubsection.
\begin{lstlisting}
$ python XGuardian/XGuardian/Experiments/5.6_Generalizability/farlight84/src/XGuardian/XGuardian-Inspector.py XGuardian/XGuardian/Experiments/5.6_Generalizability/farlight84/dataset/mobile/Farlight84_Mobile_train.csv XGuardian/XGuardian/Experiments/5.6_Generalizability/farlight84/dataset/mobile/Farlight84_Mobile_val.csv XGuardian/XGuardian/Experiments/5.6_Generalizability/farlight84/dataset/mobile/Farlight84_Mobile_test.csv <output_dir> 6 1 "isCheater,steamid,kill_num,match_id,angle,x,y,platform" 
$ python XGuardian/XGuardian/Design/src/XGuardian-Inspector-Match-Preprocessing.py XGuardian/XGuardian/Experiments/5.6_Generalizability/farlight84/result/mobile/aggregated_predictions_train.csv XGuardian/XGuardian/Experiments/5.6_Generalizability/farlight84/result/mobile/aggregated_predictions_train_cleaned.csv
$ python XGuardian/XGuardian/Design/src/XGuardian-Inspector-Match-Preprocessing.py XGuardian/XGuardian/Experiments/5.6_Generalizability/farlight84/result/mobile/aggregated_predictions_val.csv XGuardian/XGuardian/Experiments/5.6_Generalizability/farlight84/result/mobile/aggregated_predictions_val_cleaned.csv
$ python XGuardian/XGuardian/Design/src/XGuardian-Inspector-Match-Preprocessing.py XGuardian/XGuardian/Experiments/5.6_Generalizability/farlight84/result/mobile/aggregated_predictions_test.csv XGuardian/XGuardian/Experiments/5.6_Generalizability/farlight84/result/mobile/aggregated_predictions_test_cleaned.csv
$ python XGuardian-Inspector-Match.py --train_path XGuardian/XGuardian/Experiments/5.6_Generalizability/farlight84/result/mobile/aggregated_predictions_train_cleaned.csv --val_path XGuardian/XGuardian/Experiments/5.6_Generalizability/farlight84/result/mobile/aggregated_predictions_val_cleaned.csv --test_path XGuardian/XGuardian/Experiments/5.6_Generalizability/farlight84/result/mobile/aggregated_predictions_test_cleaned.csv --feature_set 'o' --output_dir <output_directory>
\end{lstlisting}

\subsubsection{Case Study}
\textbf{[\textit{To check the raw results}]} Each subfolder of \texttt{XGuardian\allowbreak/XGuardian\allowbreak/Experiments\allowbreak/5.7\allowbreak\_Case\allowbreak\_Study} contains the corresponding results.

\noindent\textbf{[\textit{To reproduce}]} For elimination scope, please follow the procedures in \autoref{sec:explainer}. For match scope, run the following command.
\begin{lstlisting}
$ python XGuardian/XGuardian/Experiments/5.7_Case_Study/match_scope/shap_case_Study.py --train_path XGuardian/XGuardian/Experiments/5.2_Detection_Performace/overall_performance/TrainValTest/aggregated_predictions_train_cleaned.csv --dataset_path XGuardian/XGuardian/Experiments/5.2_Detection_Performace/overall_performance/TrainValTest/aggregated_predictions_test_cleaned.csv --model_path XGuardian/XGuardian/Experiments/5.2_Detection_Performace/overall_performance/TrainValTest/logistic_regression_model_original.joblib --feature_set 'o' --output_dir <output_dir>
\end{lstlisting}

\subsubsection{Robustness to Adversarial Attack}
\textbf{[\textit{To check the raw results}]} Use \texttt{XGuardian\allowbreak/XGuardian\allowbreak/Experiments\allowbreak/5.8\allowbreak\_Adversarial\allowbreak\_Attack\allowbreak/src\allowbreak/round\allowbreak\_eval\allowbreak\_metrics\allowbreak\_stats.py} to calculate the metrics. The directory of the file \texttt{round\allowbreak\_integ\allowbreak\_test\allowbreak\_predictions\allowbreak\_original.csv} must be modified before the execution at Line 5.
\begin{lstlisting}
$ python XGuardian/XGuardian/Experiments/5.8_Adversarial_Attack/src/round_eval_metrics_stats.py
\end{lstlisting}

\noindent\textbf{[\textit{To reproduce}]} Run the following commands. Due to minimizing the size of our artifacts, we deleted some intermediate files (e.g., \texttt{aggregated\allowbreak\_prediction\allowbreak\_x.csv}, etc.), you may reproduce it by using \texttt{XGuardian\allowbreak/XGuardian\allowbreak/Experiments\allowbreak/5.8\allowbreak\_Adversarial\allowbreak\_Attack\allowbreak/src\allowbreak/XGuardian\allowbreak-Inspector\allowbreak-Prediction\allowbreak-only.py}. Note that the directories of the models \texttt{XGuardian\allowbreak/XGuardian\allowbreak/Experiments\allowbreak/5.2\allowbreak\_Detection\allowbreak\_Performace\allowbreak/overall\allowbreak\_performance\allowbreak/TrainValTest\allowbreak/gru\allowbreak\_cnn\allowbreak\_best\allowbreak\_model.keras}, \texttt{XGuardian\allowbreak/XGuardian\allowbreak/Experiments\allowbreak/5.2\allowbreak\_Detection\allowbreak\_Performace\allowbreak/overall\allowbreak\_performance\allowbreak/TrainValTest\allowbreak/aggregation\allowbreak\_best\allowbreak\_model.keras} and \texttt{XGuardian\allowbreak/XGuardian\allowbreak/Experiments\allowbreak/5.2\allowbreak\_Detection\allowbreak\_Performace\allowbreak/overall\allowbreak\_performance\allowbreak/TrainValTest\allowbreak/best\allowbreak\_aggregation\allowbreak\_threshold\allowbreak\_with\allowbreak\_attention.txt} should be properly modified at Line 187-190, Line 256-260 and Line 263-268, respectively.
Simply, you may instead copy these models to the \texttt{<output\allowbreak\_dir>} directly.
\begin{lstlisting}
$ python XGuardian/XGuardian/Experiments/5.8_Adversarial_Attack/src/XGuardian-Inspector-Prediction-only.py XGuardian/XGuardian/Experiments/5.8_Adversarial_Attack/data/adversarial_dataset.csv XGuardian/XGuardian/Experiments/5.8_Adversarial_Attack/result 6 1 "isCheater,steamid,kill_num,match_id,in_crouch,is_walking,is_strafing,weapon,angle,x,y,distance" 
$ python XGuardian/XGuardian/Experiments/5.8_Adversarial_Attack/src/XGuardian-Inspector-Match-Preprocessing.py XGuardian/XGuardian/Experiments/5.8_Adversarial_Attack/result/aggregated_predictions_test.csv XGuardian/XGuardian/Experiments/5.8_Adversarial_Attack/result/aggregated_predictions_test_cleaned.csv 
$ python XGuardian/XGuardian/Experiments/5.8_Adversarial_Attack/src/XGuardian-Inspector-Match-Prediction-only.py --test_path XGuardian/XGuardian/Experiments/5.8_Adversarial_Attack/result/aggregated_predictions_test_cleaned.csv --feature_set 'o' --model_path XGuardian/XGuardian/Experiments/5.8_Adversarial_Attack/result/logistic_regression_model_original.joblib --output_dir XGuardian/XGuardian/Experiments/5.8_Adversarial_Attack/result
\end{lstlisting}

\noindent\textbf{[\textit{Reuse of \inspector without training}]} Please note that the above three scripts can be used to reuse the pre-trained model from \sys, solely for performing prediction functions.

\subsubsection{Overhead}
\textbf{[\textit{To check the raw results}]} Please check \texttt{data\allowbreak preprocessing\allowbreak\_overhead.log}, \texttt{initial\allowbreak model\allowbreak training\allowbreak\_overhead.log} and \texttt{testing\allowbreak\_overhead.log} at \texttt{XGuardian\allowbreak/XGuardian\allowbreak/Experiments\allowbreak/5.9\allowbreak\_Overhead\allowbreak/overhead}.

\noindent\textbf{[\textit{To reproduce}]} Run the following command with \texttt{detector\allowbreak\_cleaned} environment.
\begin{lstlisting}
$ python XGuardian/XGuardian/Experiments/5.9_Overhead/overhead/GRU_aggregated_2D_feature_selection.py XGuardian/XGuardian/Experiments/5.9_Overhead/overhead/Train_cleaned.csv XGuardian/XGuardian/Experiments/5.9_Overhead/overhead/Validation_cleaned.csv XGuardian/XGuardian/Experiments/5.9_Overhead/overhead/Test_cleaned.csv <output_dir> 6 1 "isCheater,steamid,kill_num,match_id,in_crouch,is_walking,is_strafing,weapon,angle,x,y,distance"
\end{lstlisting}

\subsubsection{User Study}
The original demos are stored at \texttt{XGuardian\allowbreak/XGuardian\allowbreak/Experiments\allowbreak/5.10\allowbreak\_User\allowbreak\_Study\allowbreak\_(addtional\allowbreak\_evaluation)}. Please check our paper for the answer of the questionnaires and the operation procedures.




\subsection{Version}
Based on the LaTeX template for Artifact Evaluation V20231005. Submission,
reviewing and badging methodology followed for the evaluation of this artifact
can be found at \url{https://secartifacts.github.io/usenixsec2026/}.
